\pdfoutput=1 
\documentclass[fleqn,usenatbib,letterpaper]{mnrasletter}
\usepackage[totalwidth=480pt,totalheight=680pt,layoutvoffset=0.5cm]{geometry}

\usepackage{newtxtext,newtxmath}

\usepackage[T1]{fontenc}
\usepackage{ae,aecompl}

\usepackage{graphicx}	
\usepackage{amsmath}	
\usepackage{amssymb}	
\usepackage{breqn}

\title[Secular resonance sculpting in debris discs]{Empty gaps? Depleting annular regions in debris discs by secular resonance with a two-planet system}

\author[B. Yelverton \& G. M. Kennedy]{
Ben Yelverton$^{1}$\thanks{E-mail: bmy21@cam.ac.uk} and
Grant M. Kennedy$^{2, 3}$
\\
$^{1}$Institute of Astronomy, University of Cambridge, Madingley Road, Cambridge CB3 0HA, UK\\
$^{2}$Department of Physics, University of Warwick, Gibbet Hill Road, Coventry CV4 7AL, UK\\
$^{3}$Centre for Exoplanets and Habitability, University of Warwick, Gibbet Hill Road, Coventry CV4 7AL, UK\\
}

\date{Accepted XXX. Received YYY; in original form ZZZ}

\pubyear{2018}

\begin{document}
\label{firstpage}
\pagerange{\pageref{firstpage}--\pageref{lastpage}}
\maketitle

\begin{abstract}
We investigate the evolution on secular time-scales of a radially extended debris disc under the influence of a system of two coplanar planets interior to the disc, showing that the secular resonances of the system can produce a depleted region in the disc by exciting the eccentricities of planetesimals. Using Laplace-Lagrange theory, we consider how the two exterior secular resonance locations, time-scales and widths depend on the masses, semi-major axes and eccentricities of the planets. In particular, we find that unless the resonances are very close to each other, one of them is very narrow and therefore unimportant for determining the observable structure of the disc. We apply these considerations to the debris disc of HD~107146, identifying combinations of the parameters of a possible unobserved two-planet system that could configure the secular resonances appropriately to reproduce the depletion observed in the disc. By performing \textit{N}-body simulations of such systems, we find that planetesimal eccentricities do indeed become large near the theoretical secular resonance locations. The \textit{N}-body output is post-processed to set the initial surface density profile of the disc, and to include the possible effects of collisional depletion. We find that it is possible to obtain a double-ringed disc in these simulations but not an axisymmetric one, with the inner ring having an offset whose magnitude depends on the eccentricities of the planets, and the outer ring showing spiral structure.

\end{abstract}




\section{Introduction}

Planetary systems in general contain not only planets, but also a range of other objects ranging from $\sim$$\mu$m-sized particles up to planetesimals of diameter $\sim$1000km (\citealt{Wyatt08_Review}); these are the constituents of debris discs. The population of objects of a given size in a debris disc is replenished by the products of destructive collisions between larger objects. In this way, mass is passed from the largest objects down to smaller objects in a collisional cascade (e.g. \citealt{Backman1993_Circumstellar}; \citealt{Dominik2003_CC}). The disc loses mass over time, as the particles eventually become so small that they are blown out of the system by radiation pressure. Though it is difficult to probe the population of large planetesimals directly, observations at $\sim$mm wavelengths trace the distribution of these parent planetesimals, since $\sim$mm-sized grains are not strongly affected by radiation forces.

The planetesimals comprising debris discs are dynamically perturbed by planets, and these perturbations can heavily influence their spatial distribution. Such planet-disc interactions are evident in our own solar system, for example in the Kirkwood gaps in the asteroid belt, which are in mean motion resonance with Jupiter (e.g. \citealt{Murray1998}). Studying the structure of a debris disc can thus be a powerful way to learn about planets that may be present. For example, a warp in the disc of $\beta$ Pictoris led to the prediction of an inclined planet interior to the disc (\citealt{Mouillet97_betaPicPred}); such a planet was later confirmed by direct imaging (\citealt{Lagrange2010_betaPicImage}). 

This paper proposes a mechanism for the formation of depleted regions -- i.e. `gaps' -- in debris discs via dynamical interactions with a planetary system, with a particular focus on the disc around HD~107146, a $\sim$100Myr old Sun-like star (\citealt{Williams04_Age}). \citet{Ricci15_AlmaObs} observed this disc using the Atacama Large Millimeter/submillimeter Array (ALMA), and found that it has an unusual structure: the radial surface density profiles that best fit the data have inner and outer edges of $\sim$30au and $\sim$150au, and a depletion at $\sim$70--80au. More recent high resolution ALMA observations have confirmed this depleted structure (\citealt{Marino2018_107146}).

\citet{Ricci15_AlmaObs} suggested that the depletion could be the result of a planet of several Earth masses on a near-circular orbit at $\sim$80au ejecting nearby planetesimals from the system. However, this raises the question of why only a single planet formed in such a broad disc, and why it did so at such a large distance from the star. In fact, \citet{Marino2018_107146} conclude based on their updated data that any planets in the depleted region would need to be tens of Earth masses, which would be even more difficult to form. Therefore, alternative scenarios should be considered, in which the depletion is the result of interactions with planets closer to the star. For example, \citet{Pearce15_EccPlanet} used $N$-body simulations to demonstrate that an eccentric planet of comparable mass to the disc (which they took to be $100M_{\oplus}$) with semi-major axis $\sim$30--40au could create a dip in the surface density at the observed location through its interactions with the disc. 

In this paper, we will consider a scenario in which there are multiple planets interior to the disc (i.e. within 30au), sculpting the disc via their secular resonances -- these resonances are important for understanding the distribution of asteroids in our own asteroid belt (e.g. \citealt{Milani90_SecRes}), so it is natural to seek to use them to understand other debris discs. At the secular resonances, the rate of pericentre precession of the planetesimals (induced by the planets) is equal to one of the eigenfrequencies of the system (the characteristic frequencies that determine the rates of precession of the planets themselves), which leads to large planetesimal eccentricities (e.g. \citealt{Murray1998}). Eccentric planetesimals move over a wider range of radii than those on near-circular orbits, so on purely dynamical grounds we expect to see a depletion in surface density at the radial locations of the secular resonances even if the planetesimals there are not ejected. Increasing the eccentricity of planetesimals will also enhance the rate of catastrophic collisions, which will further deplete the disc around the resonances.

The reason for requiring multiple planets to be present in this scenario is that for there to be any secular resonances in a planetary system, the planets must be precessing, which will always be the case if there are at least two of them. \citet{Zheng2017_GasSR} showed that a single planet can deplete a debris disc via secular resonance sweeping if there is a gas disc present causing it to precess, but in that scenario the depleted area is a very broad region in which the planet is embedded. Multi-planet systems are not uncommon, with 627 of them currently known\footnote{From the Extrasolar Planets Encycopaedia (\citealt{Schneider2011_Cat})}. Many of these also contain debris discs -- for example, HR~8799 hosts four giant planets at tens of au (\citealt{Marois2008_8799}; \citealt{Marois2010_8799}) and an extended disc (e.g. \citealt{Booth2016_8799}), an architecture that is broadly similar to those we will consider in this paper. A system that is more representative of the currently known multi-planet systems is 61~Vir, which also has an extended debris disc, and three planets with masses of order 10$M_\oplus$ lying within 1au (\citealt{Wyatt2012_61Vir}; \citealt{Marino2017_61Vir}). If a broad debris disc covering an appropriate range of distances from the star is present in a multi-planet system, it will inevitably be sculpted by secular resonances, naturally giving rise to depleted regions.

The depleted structure of HD~107146 is not unique among debris discs. Scattered light observations of HD~92945 suggested that the disc in this system also has two peaks in its radial density profile (\citealt{Golimowski2011_92945}), though such observations probe small particles which will be affected by radiation forces. Recent ALMA observations have shown that the $\sim$mm-sized dust in this system also has a depleted profile (Marino et al. in preparation), which implies that this is also the case for the larger planetesimals in the disc. In addition, HD~131835 is host to a \textit{triple}-peaked disc (\citealt{Feldt2016_131835}), though this disc is known to contain CO gas (\citealt{Moor2015_Gas}), which complicates the dynamics of dust in this system. 

\citet{MoroMartin08_38529} applied the theory of secular resonances to the two known planetary companions of the star HD~38529, and from this constrained the location of the planetesimals responsible for the dust emission from its unresolved debris disc, concluding that they are located between the resonances. In this paper we aim to perform what is essentially the reverse of this procedure. HD~107146 has no known planets, though in the context of the planets invoked in this paper the limits are not particularly stringent. Given the structure of its disc we will place constraints on the combinations of masses and semi-major axes of a two-planet system that would place secular resonances appropriately to generate a depletion at the observed location. We will then perform numerical simulations of the dynamical and collisional evolution of the disc using parameters we identified as suitable, constructing synthetic images and radial intensity profiles to compare with observations. We are able to make a more detailed comparison between simulations and observations than in \citet{MoroMartin08_38529}, since we now have high-resolution ALMA images rather than only spectral energy distributions.

In the following section, we give an outline of the theoretically expected secular evolution of the disc, considering in particular how the characteristics of the secular resonances depend on the nature of the planets. In section \ref{sec:PS_constraints}, we apply these considerations to HD~107146 in order to make deductions about what kind of two-planet system could deplete the disc in the manner observed. Section~\ref{sec:sims} describes the methods we use to numerically simulate the influence of a system of planets on the disc, and presents the results of simulations of some promising configurations. In section~\ref{sec:discussion} we first address some issues with the model as applied to HD~107146, before discussing some other systems where the secular resonance depletion model may be relevant. Section~\ref{sec:conclusions} then presents the overall conclusions of the work.

\section{Secular Resonances}
\label{sec:SR_theory}

We wish to understand the long-term interaction between a system of planets and a debris disc; since the latter is composed of comparatively low-mass planetesimals, we treat it as a collection of massless test particles. The effect of this interaction can be calculated analytically by considering the disturbing function, which quantifies the dynamical perturbations from the planets on each other and on a test particle in the disc. It is usual to expand the disturbing function as an infinite series in the orbital elements of the planets and the test particle, as for a given problem only certain terms will be relevant.

The terms of interest to us here are the \textit{secular} terms, those that do not depend on how far the orbiting bodies are along their orbits, and therefore give the evolution of the system on time-scales much greater than an orbital period. By considering these terms up to second order in eccentricities and inclinations, one can derive the results of Laplace-Lagrange theory (\citealt{Murray1998}), which we outline below. 

This theory can be applied to a system containing an arbitrary number of planets $N_{\mathrm{pl}}$, but here we restrict our attention to $N_{\mathrm{pl}}=2$, even though a system of three or more planets could work equally well. This is because ultimately we aim to identify regions in the space of planetary parameters that can give a disc structure matching observations; the problem is not a well constrained one, so we will consider only the case that minimises the number of free parameters. 

\subsection{Laplace-Lagrange Theory}
\label{sec:LL_theory} 

Consider a system of two planets with masses $m_j$, semi-major axes $a_j$ and eccentricities $e_j$, where $j=1$ refers to the innermost planet. The mean motion of planet $j$ is $n_j \approx \sqrt{G m_{\mathrm{c}}/ a_j^3}$, where $m_{\mathrm{c}}$ is the mass of the central star. We are interested in the evolution of a test particle's eccentricity $e$ and longitude of pericentre $\varpi$. The test particle's semi-major axis (which remains constant in this theory) is $a$, and its mean motion is $n \approx \sqrt{G m_{\mathrm{c}}/ a^3}$.

The evolution of $e$ and $\varpi$ is described in terms of the eccentricity vector $(k,h)$, defined such that its magnitude is $e$ and its phase (i.e. angle relative to the $k$-axis) is $\varpi$:

\begin{equation}\label{eqn:hkdef}
\begin{split}
	k &= e\cos\varpi, \\
	h &= e\sin\varpi.
\end{split}
\end{equation}

To give the solution for these variables, we must first define some intermediate quantities. Consider the $2\times 2$  matrix $\mathbf{A}$, which has entries

\begin{equation}\label{eqn:Ajj}
	A_{jj}=\frac{n_j}{4}\frac{m_k}{m_{\mathrm{c}}}\alpha_{12}\bar{\alpha}_{12}b_{3/2}^{(1)}(\alpha_{12}),
\end{equation}

\begin{equation}\label{eqn:Ajk}
	A_{jk}=-\frac{n_j}{4}\frac{m_k}{m_{\mathrm{c}}}\alpha_{12}\bar{\alpha}_{12}b_{3/2}^{(2)}(\alpha_{12}),
\end{equation}

where $j\neq k$, the $b_s^{(\lambda)}$ are Laplace coefficients, $\alpha_{12}=\frac{a_1}{a_2}$, $\bar{\alpha}_{12}=\alpha_{12}$ if $j=1$ and $\bar{\alpha}_{12}=1$ if $j=2$. We denote the eigenvalues, or eigenfrequencies, of this matrix by $g_i$, and the $j^{\mathrm{th}}$ component of the eigenvector corresponding to $g_i$ by $e_{ji}$. We define also the quantities

\begin{equation}\label{eqn:Aj}
	A_j=-\frac{n}{4}\frac{m_j}{m_{\mathrm{c}}}\alpha_j\bar{\alpha}_jb_{3/2}^{(2)}(\alpha_j)
\end{equation}

\noindent and 

\begin{equation}\label{eqn:Adef}
	A=\frac{n}{4}\sum_{j=1}^{2}\frac{m_j}{m_{\mathrm{c}}}\alpha_j\bar{\alpha}_jb_{3/2}^{(1)}(\alpha_j),
\end{equation}

\noindent where

\begin{equation}\label{eqn:alphaj}
	\alpha_j=
	\begin{cases}
		a_j/a & \text{if}\ a_j<a, \\
		a/a_j & \text{if}\ a_j>a,
	\end{cases}
\end{equation}

\begin{equation}\label{eqn:alphabarj}
	\bar{\alpha}_j=
	\begin{cases}
		1 & \text{if}\ a_j<a, \\
		a/a_j & \text{if}\ a_j>a.
	\end{cases}
\end{equation}

\noindent The solution for the eccentricity vector is then

\begin{equation}\label{eqn:hksol}
\begin{split}
	k&=e_{\mathrm{free}}\cos{(At+\beta)} + k_0(t), \\
	h&=e_{\mathrm{free}}\sin{(At+\beta)} + h_0(t) .
\end{split}
\end{equation}

Here, $e_{\mathrm{free}}$ and $\beta$ are constants set by the initial conditions of the problem; $k_0$ and $h_0$ are defined as

\begin{equation}\label{eqn:h0k0}
\begin{split}
	k_0(t)&=-\sum_{i=1}^{2}\frac{\nu_i}{A-g_i}\cos{(g_it+\beta_i}), \\
	h_0(t)&=-\sum_{i=1}^{2}\frac{\nu_i}{A-g_i}\sin{(g_it+\beta_i)}, 
\end{split}
\end{equation}

\noindent where 

\begin{equation}\label{eqn:nui}
	\nu_i=\sum_{j=1}^{2}A_je_{ji},
\end{equation}

and $\beta_i$ and the scaling of the eigenvectors $e_{ji}$ are also set by the initial conditions. Equation~(\ref{eqn:hksol}) shows that the eccentricity vector is the sum of two other vectors -- the \textit{free} component, with magnitude $e_{\mathrm{free}}$ and phase $\varpi_{\mathrm{free}}=At+\beta$, and the \textit{forced} component, whose magnitude is $e_{\mathrm{forced}}=\sqrt{k_0^2+h_0^2}$ and phase $\varpi_{\mathrm{forced}}=\arctan{\left(\frac{h_0}{k_0}\right)}$. 

At any given semi-major axis, the free component of the eccentricity vector has constant magnitude, and its argument is increasing linearly with time at a rate $\dot{\varpi}_{\mathrm{free}}=A$. That is, the free component of the longitude of pericentre is precessing at a constant rate, as would the pericentre of a particle in a single-planet system. In the $(k,h)$ plane, this corresponds to anticlockwise circular motion. The instantaneous centre of the circle is the forced eccentricity vector $(k_0,h_0)$, which itself is varying in time in a non-trivial manner. So, the evolution of the total eccentricity vector in the $(k,h)$ plane is that of a point which at each instant is moving in a circle around the (moving) point specified by equation~(\ref{eqn:h0k0}).

It is also clear from equation~(\ref{eqn:h0k0}) that $e_{\mathrm{forced}}$ reaches very large values where the precession frequency $A(a)$ is close to one of the eigenvalues $g_i$; the locations where this condition is satisfied are the secular resonances. In this theory, the eccentricity can reach arbitrarily high values, which is unphysical. Including higher order terms in the disturbing function would limit the growth of the eccentricity -- this is discussed by \citet{Malhotra98_Resonances} using the Hamiltonian formalism, for the simpler case of a single planet whose pericentre is artificially made to precess. However, this would make it much more difficult to obtain an analytical solution to the secular problem. 

The time evolution of the planetary orbits can also be determined within Laplace-Lagrange theory. We define the quantities

\begin{equation}\label{eqn:hjkj}
\begin{split}
	k_j(t)&=\sum_{i=1}^{2}e_{ji}\cos{(g_it+\beta_i}), \\
	h_j(t)&=\sum_{i=1}^{2}e_{ji}\sin{(g_it+\beta_i)}. 
\end{split}
\end{equation}

Then, the eccentricities of the planets are given by $e_{j}=\sqrt{k_j^2+h_j^2}$ and their longitudes of pericentre by $\varpi_j=\arctan{\left(\frac{h_j}{k_j}\right)}$. The latter of these relations highlights the fact that in general it is not simply the case that the pericentre of the inner planet precesses at frequency $g_1$ and the outer at $g_2$. Rather, both eigenfrequencies contribute to both rates of precession $\dot{\varpi}_j$. Depending on the planetary semi-major axes and masses in a way that will be considered in the following subsection, $\dot{\varpi}_1$ can approximate to either $g_1$ or $g_2$, and the same is true of $\dot{\varpi}_2$.

In the above discussion, no mention was made of the inclination $I$ of the test particle. It can be shown that in general $I$ evolves in a way that is analogous to equation~(\ref{eqn:hksol}), reaching large values at semi-major axes where the rate of precession of the longitude of ascending node $\Omega$ is equal to one of the inclination eigenfrequencies (which are not the same as $g_i$). However, we will consider only planets with zero inclination; in this special case the forced inclination remains zero everywhere and the inclination resonances have no effect on the disc. Even if the planets were inclined, the effect of the inclination resonances would not necessarily be clear observationally, as it is much more difficult to observe a radially dependent scale height than a radially dependent surface density. This is especially true for HD~107146, the disc that we take as an example in sections \ref{sec:PS_constraints} and \ref{sec:sims}, as its orientation is nearly face-on. So, in the remainder of this paper, the term `secular resonance' always refers to a secular \textit{eccentricity} resonance.

\subsection{Resonance Locations and Time-scales}
\label{sec:sr_locations} 

We now consider how the properties of the secular resonances -- firstly, their locations and the time-scales on which they excite eccentricities -- depend on the properties of the planets. To do this, consider first the eigenfrequencies $g_i$ of the system; using equations (\ref{eqn:Ajj}) and (\ref{eqn:Ajk}), these can be written as

\begin{dmath}\label{eqn:gi}
	g_{1,2}=\frac{n_1}{4}{\frac{m_1}{m_{\mathrm{c}}}\left(\frac{a_1}{a_2}\right)}^2\left[\frac{1}{2}b^{(1)}_{3/2}\left(\frac{a_1}{a_2}\right)\left(\frac{m_2}{m_1}+\sqrt{\frac{a_1}{a_2}}\right)\pm\sqrt{\frac{1}{4}\left[b^{(1)}_{3/2}\left(\frac{a_1}{a_2}\right)\right]^2\left(\frac{m_2}{m_1}-\sqrt{\frac{a_1}{a_2}}\right)^2+\sqrt{\frac{a_1}{a_2}}\frac{m_2}{m_1}\left[b^{(2)}_{3/2}\left(\frac{a_1}{a_2}\right)\right]^2}\right].
\end{dmath}

The condition for secular resonance is $A(a)=g_i$. Since part of the motivation for this paper is to remove the need for a planet embedded in a depletion at a large distance from the star, instead having closer-in planets produce the depletion from a distance, we will consider only the resonances exterior to the planets. Since $A(a)$ is monotonically decreasing for $a>a_2$, there are two relevant resonances, whose semi-major axes are, from equations (\ref{eqn:Adef}) and (\ref{eqn:gi}), the solutions of

\begin{dmath}\label{eqn:sr_condition}
\frac{1}{2}b^{(1)}_{3/2}\left(\frac{a_1}{a_2}\right)\left(\frac{m_2}{m_1}+\sqrt{\frac{a_1}{a_2}}\right)\pm\sqrt{\frac{1}{4}\left[b^{(1)}_{3/2}\left(\frac{a_1}{a_2}\right)\right]^2\left(\frac{m_2}{m_1}-\sqrt{\frac{a_1}{a_2}}\right)^2+\sqrt{\frac{a_1}{a_2}}\frac{m_2}{m_1}\left[b^{(2)}_{3/2}\left(\frac{a_1}{a_2}\right)\right]^2}\\={\left(\frac{a_1}{a_2}\right)}^{1/2}\left[b^{(1)}_{3/2}\left(\frac{a_1}{a}\right)+\frac{m_2}{m_1}\frac{a_2}{a_1}b^{(1)}_{3/2}\left(\frac{a_2}{a}\right)\right].
\end{dmath}

For the case with the positive sign, we denote the solution for $a$ by $r_1$, as this is the resonance corresponding to the faster eigenfrequency $g_1$. The solution in the case of the negative sign, corresponding to $g_2$, is denoted by $r_2$, and we have $r_1<r_2$.

\begin{figure}
	\centering
    \hspace{-0.5cm}
	\includegraphics[width=0.5\textwidth]{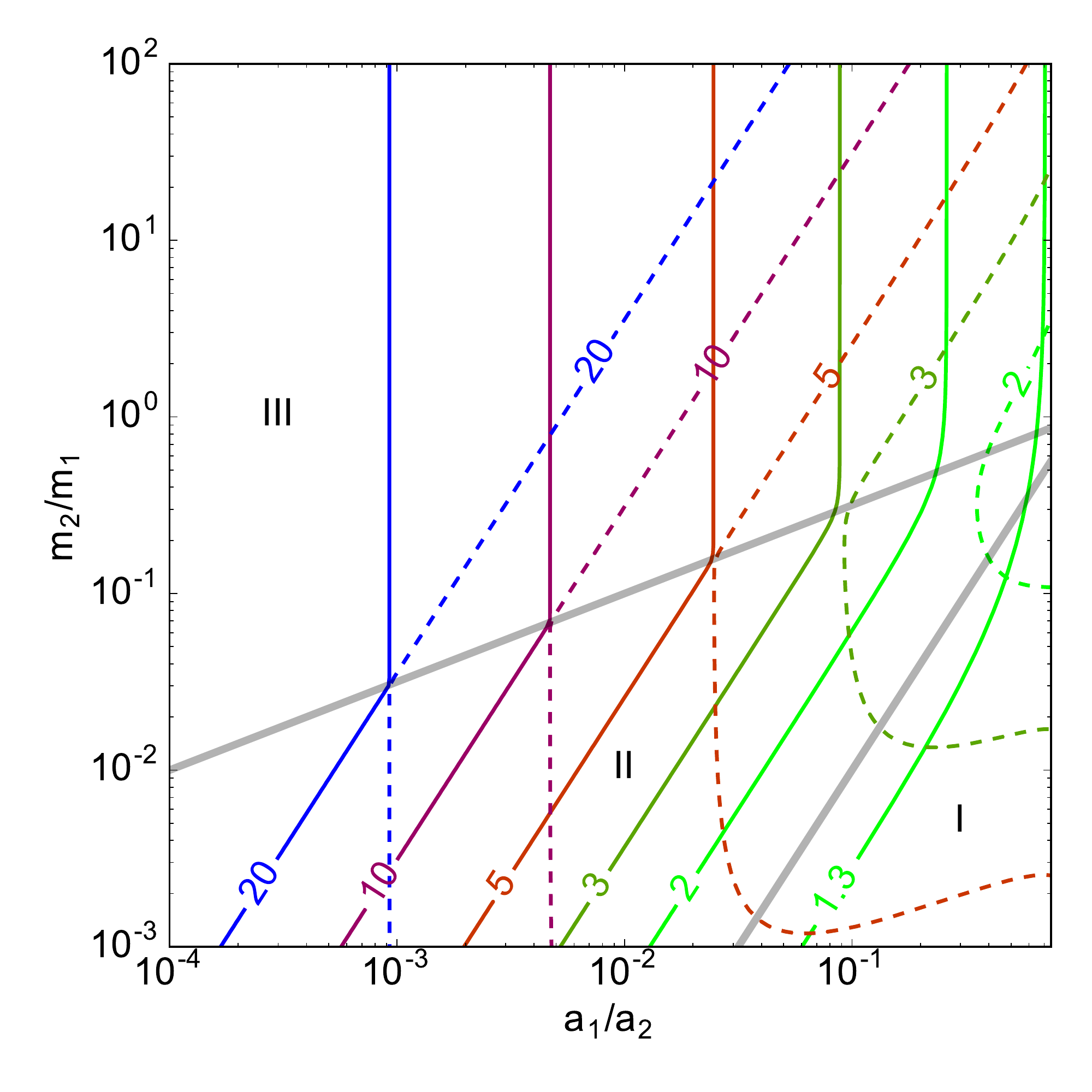}
	\caption{Contour plot showing the exterior secular resonance locations of a two-planet system relative to the outermost planet, in the space of ratios of planetary semi-major axes and masses. The solid coloured lines show constant $\frac{r_1}{a_2}$ and the dashed lines $\frac{r_2}{a_2}$. The thick grey lines are $y=x^2$ and $y=\sqrt{x}$, which separate the regions I, II and III, in which the contours have different characteristic slopes. The lines that separate the three regions extend to arbitrarily low values of $\frac{m_2}{m_1}$ and $\frac{a_1}{a_2}$; the dashed contours with values 10 and 20 have the same form as the others, but turn over at lower mass ratios than are shown on the plot. The closer the planets are to each other, the closer the secular resonances are to the outermost planet.}
	\label{fig:srloc}
\end{figure}

\begin{figure}
	\centering
    \hspace{-0.5cm}
	\includegraphics[width=0.5\textwidth]{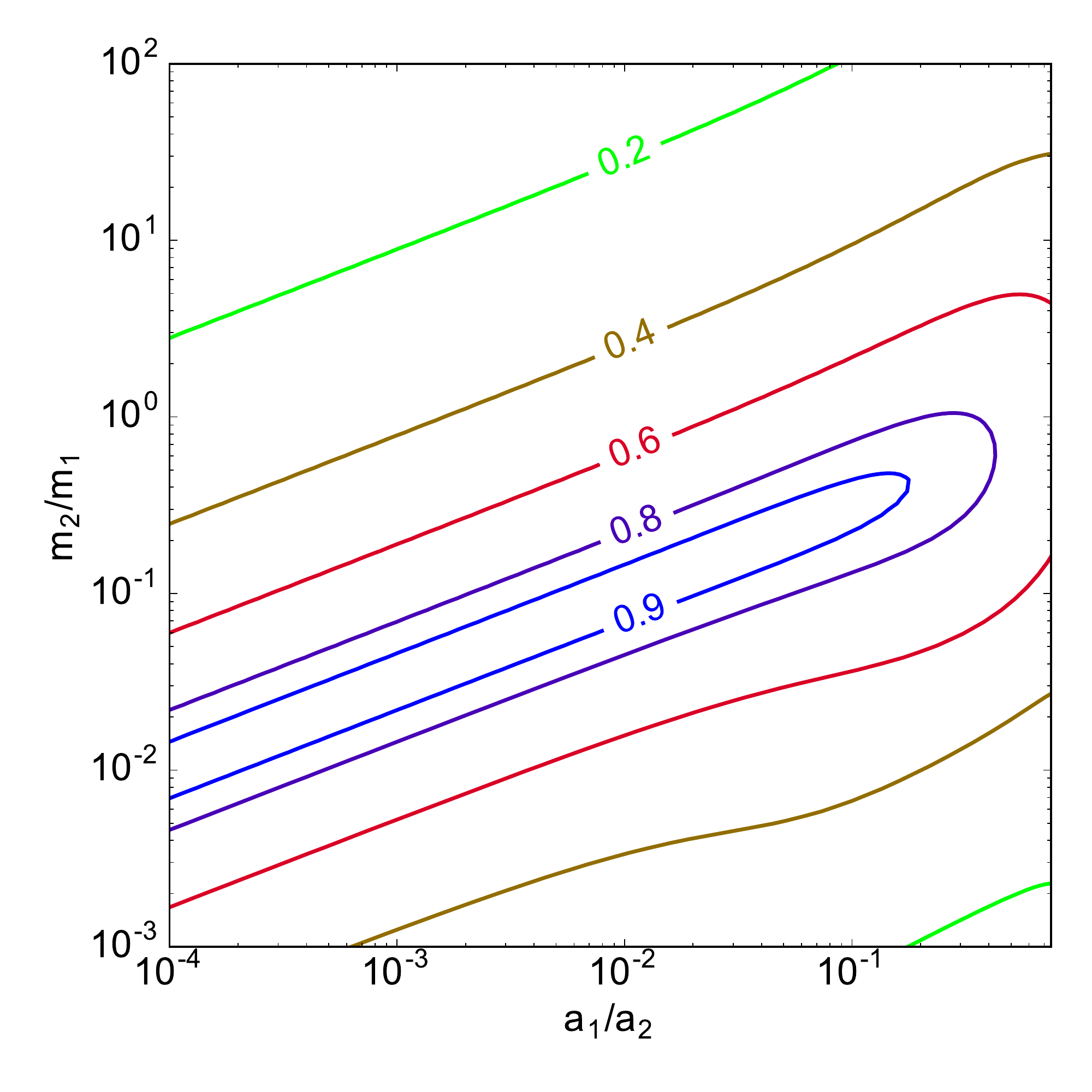}
	\caption{Contour plot showing the ratio between the exterior secular resonance locations, $\frac{r_1}{r_2}$, of a two-planet system, in the space of ratios of planetary semi-major axes and masses. The resonances are further apart for more extreme (specifically, far from the line $y=\sqrt{x}$) mass ratios.}
	\label{fig:r1r2}
\end{figure}

In equation~(\ref{eqn:sr_condition}), the planet masses appear only in the combination $\frac{m_2}{m_1}$, and the three relevant semi-major axes appear only in the combinations $\frac{a_1}{a_2}$, $\frac{a_1}{a}$ and $\frac{a_2}{a}$. Because $\frac{a_1}{a}=\frac{a_1}{a_2}\frac{a_2}{a}$, only two of these combinations are independent, and we can solve for $\frac{a}{a_2}$ (or $\frac{a}{a_1}$) as a function of $\frac{a_1}{a_2}$ and $\frac{m_2}{m_1}$. That is, the positions of the resonances \textit{relative} to the planets can be calculated given only the \textit{ratios} of the planet parameters. This is illustrated in Fig.~\ref{fig:srloc}, which shows contours of $\frac{r_1}{a_2}$ and $\frac{r_2}{a_2}$, evaluated numerically. The interpretation of this figure will be considered in detail below. By taking the ratio of these quantities, contours of $\frac{r_1}{r_2}$ can also be calculated, as shown in Fig.~\ref{fig:r1r2}.

\begin{figure}
	\centering
    \hspace{-0.5cm}
	\includegraphics[width=0.5\textwidth]{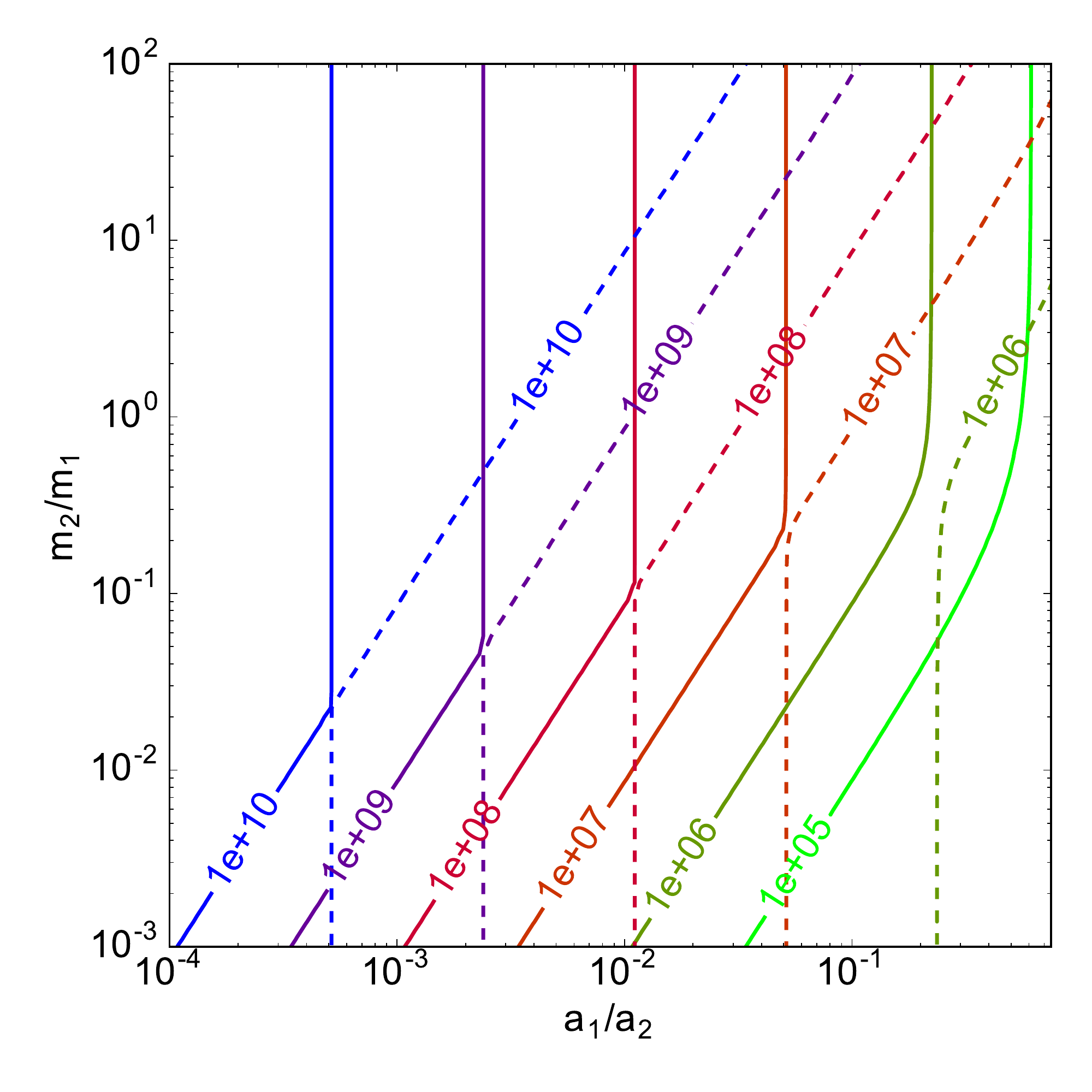}
	\caption{Contour plot showing the secular precession time-scales at the exterior secular resonances of a two-planet system, $\tau_1$/yr (solid lines) and $\tau_2$/yr (dashed lines), in the space of ratios of planetary semi-major axes and masses. Here, the outermost planet has parameters $a_2=25\mathrm{au}$ and $m_2=1.5M_{\mathrm{J}}$. The time-scales are shorter if the planets are closer together (but the resonances are also closer in, from Fig.~\ref{fig:srloc}).} 
	\label{fig:time-scale}
\end{figure}

We can also quantify the time-scales over which eccentricities are excited at the secular resonances. The motion of particles in $(e\cos\varpi,e\sin\varpi)$ space is complicated, but we can expect to see a significant change in the eccentricity at the resonances on a time-scale equal to the period of secular precession there, i.e. the time-scale on which particles circulate in that space. From section~\ref{sec:LL_theory}, the frequency of this precession at a distance of $r_i$ is $A(r_i)$, which by definition of $r_i$ is equal to $g_i$; thus, the relevant time-scale at the $i^{\mathrm{th}}$ resonance is given by $\tau_i=2\pi/g_i$. The $N$-body simulations we perform in section~\ref{sec:sims} verify that this is indeed a reasonable estimate of the excitation time-scale. Equation~(\ref{eqn:gi}) can be used to plot contours of constant $\tau_i$, as shown in Fig.~\ref{fig:time-scale}. The time-scales depend not only on the ratios of planet parameters, but on their absolute values; Fig.~\ref{fig:time-scale} corresponds to a particular choice of $a_2$ and $m_2$. 

From equation~(\ref{eqn:gi}), we can write $g_i\propto\frac{m_1}{a_1^{3/2}}F_1(\frac{a_1}{a_2},\frac{m_2}{m_1})$, or equivalently $g_i\propto\frac{m_2}{a_2^{3/2}}F_2(\frac{a_1}{a_2},\frac{m_2}{m_1})$, where $F_1$ and $F_2$ are functions depending only on the ratios. It follows that increasing the mass, or decreasing the semi-major axis, of either planet while keeping $\frac{a_1}{a_2}$ and $\frac{m_2}{m_1}$ fixed will shorten the time-scales (though note that decreasing one of the semi-major axes also moves the resonances inwards). This makes intuitive sense -- planets secularly perturb the disc more quickly if they are more massive, or closer to the star so that their orbital time-scales are shorter. In terms of Fig.~\ref{fig:time-scale}, this means that decreasing $a_2$ or increasing $m_2$ will shift the contours to the left.

The contours in Figs. \ref{fig:srloc} -- \ref{fig:time-scale} have simple forms, being composed of straight-line segments in log space. An understanding of why this is can be gained by looking at approximate forms of $g_i$ and $A(a)$. First, we assume that $\frac{a_1}{a_2}$, $\frac{a_1}{a}$ and $\frac{a_2}{a}$ are small, so that we can approximate the Laplace coefficients in equations (\ref{eqn:Adef}) and (\ref{eqn:gi}) using

\begin{equation}\label{eqn:laplace_app}
	b^{(1)}_{3/2}(\epsilon)\approx 3\epsilon$, \quad $b^{(2)}_{3/2}(\epsilon)\approx \frac{15}{4}\epsilon^2,
\end{equation}

where $\epsilon \ll 1$ (\citealt{Murray1998}). Further simplifications can be made by considering the limits of $g_i$ and $A(a)$ in appropriate regions of parameter space. Let $x=\frac{a_1}{a_2}$ and $y=\frac{m_2}{m_1}$; after making the approximations of equation~(\ref{eqn:laplace_app}), we find that the expression for $A(a)$ contains the sum $x^2+y$, so that the appropriate limits to take are far above and below the line $y=x^2$. Similarly, inspection of equation~(\ref{eqn:gi}) shows that for $g_i$, limits should be taken above and below $y=\sqrt{x}$. The resulting limits are shown in Table~\ref{tab:limits}. Note that the limiting forms of $g_i$ immediately explain the form of Fig.~\ref{fig:time-scale}, with the curves being straight lines that switch over where $y=\sqrt{x}$. 

The parameter space is divided neatly into three regions~-- I: $y\ll x^2$, II: $x^2\ll y\ll \sqrt{x}$, and III: $y\gg \sqrt{x}$. These regions are marked on Fig.~\ref{fig:srloc}. We can find approximations for the resonance locations by equating the limiting forms of $g_i$ and $A(a)$ in each of these regions; Table~\ref{tab:approx_ratios} shows the results of doing so. Fixing $\frac{r_i}{a_2}$ then gives the slopes of the contours of constant $\frac{r_i}{a_2}$ and $\frac{r_1}{r_2}$ correctly, with one caveat: setting $g_1=A(a)$ in region I simply gives $r_1=a_2$. Fig. \ref{fig:srloc} shows that $\frac{r_1}{a_2}$ does indeed approach $1$ in that region, so this result is not incorrect -- the issue is that it is simply not possible to quantify \textit{how} the ratio approaches $1$ using the approximations made above, because the assumption $\frac{a_2}{r_1}\ll 1$ breaks down there, so that the approximations of equation~(\ref{eqn:laplace_app}) are not good.

\begin{table}
\centering
\begin{tabular}{c|c|c|c}
	\hline 
	& $g_1$ & $g_2$ & $A(a)$ \\ 
	\hline 
	$y\gg\sqrt{x}$ & $P x^{3/2}$ & $P \frac{x^2}{y}$ & - \\ 
	$y\ll\sqrt{x}$ & $P \frac{x^2}{y}$ & $P x^{3/2}$ & - \\ 
	$y\gg x^2$ & - & - & $P \left(\frac{a_2}{a}\right)^{7/2}$ \\ 
	$y\ll x^2$ & - & - & $P \left(\frac{a_2}{a}\right)^{7/2}\frac{x^2}{y}$ \\ 
    \hline 
\end{tabular}
\caption{Limiting forms of the eigenfrequencies $g_i$ and precession frequency $A(a)$, where $x=\frac{a_1}{a_2}$ and $y=\frac{m_2}{m_1}$. The pre-factor $P$ is given by $\frac{3n_2m_2}{4m_{\mathrm{c}}}$.}\label{tab:limits}
\end{table}

\begin{table}
\centering
\begin{tabular}{c|c|c|c}
	\hline 
	& $\frac{r_1}{a_2}$ & $\frac{r_2}{a_2}$ & $\frac{r_1}{r_2}$ \\ 
	\hline 
	$\mathrm{I:\ }y\ll x^2$ & 1 & $\left(\frac{x}{y^2}\right)^{1/7}$ & $\left(\frac{y^2}{x}\right)^{1/7}$ \\ 
	$\mathrm{II:\ }x^2\ll y\ll \sqrt{x}$ & $\left(\frac{y}{x^2}\right)^{2/7}$ & $x^{-3/7}$ & $\left(\frac{y^2}{x}\right)^{1/7}$ \\ 
	$\mathrm{III:\ }y\gg \sqrt{x}$ & $x^{-3/7}$ & $\left(\frac{y}{x^2}\right)^{2/7}$ & $\left(\frac{x}{y^2}\right)^{1/7}$ \\ 
    \hline 
\end{tabular}
\caption{Approximate forms of $\frac{r_i}{a_2}$ and $\frac{r_1}{r_2}$ in each region of parameter space, where $x=\frac{a_1}{a_2}$ and $y=\frac{m_2}{m_1}$.}\label{tab:approx_ratios}
\end{table}

The interpretations of Figs. \ref{fig:srloc} -- \ref{fig:time-scale} are not immediately clear, so let us take a moment to unpack the information they contain. Consider fixing the parameters of one planet -- $a_2$ and $m_2$, say -- and moving up through parameter space at fixed $\frac{a_1}{a_2}$, i.e. starting with $m_1\gg m_2$ and then decreasing $m_1$. 

Below the line $y=\sqrt{x}$, the eigenfrequency $g_2$ initially stays constant while $g_1$ decreases (which can be seen from Table~\ref{tab:limits}), and the $\tau_i$ contours of Fig.~\ref{fig:time-scale} reflect this. By taking the limit of equation~(\ref{eqn:hjkj}) for $y\ll\sqrt{x}$, it can be shown that in this region the planetary precession frequencies are $\dot{\varpi}_1\approx g_2$ and $\dot{\varpi}_2\approx g_1$. If $\frac{a_1}{a_2}$ is small enough that we start in region II (i.e. $y\gg x^2$) then, from Table~\ref{tab:limits}, the planetesimal precession frequency $A(a)$ does not depend on $\frac{a_1}{a_2}$ or $\frac{m_2}{m_1}$. Thus, the resonance at $r_2$ (which here is the resonance between the inner planet and the planetesimals) remains fixed in place, while that at $r_1$ (the resonance with the outer planet) moves outwards.

If $\frac{a_1}{a_2}$ is larger, such that we start in region I (i.e. $y\ll x^2$), then the eigenfrequencies $g_i$ behave in the same way, but from Table~\ref{tab:limits} $A(a)$ now depends on the ratios of planetary parameters, $x$ and $y$. Moving up through region I causes $A(a)$ to decrease while $g_2$ stays constant, so that $r_2$ moves inwards, until we reach region II. The fact that $A(a)$ and $g_1$ have the same dependence on $x$ and $y$ explains why $r_1$ does not vary strongly in region I.

When we move above the line $y=\sqrt{x}$, the behaviour of the eigenfrequencies switches such that $g_1$ stays constant while $g_2$ decreases; this behaviour is again evident in Fig.~\ref{fig:time-scale}. In this region $\dot{\varpi}_1\approx g_1$ and $\dot{\varpi}_2\approx g_2$. Thus, in the upper part of parameter space the resonance at $r_1$ (which is now associated with the inner planet) remains fixed, while that at $r_2$ (now associated with the outer planet) moves out.

Fig.~\ref{fig:r1r2} gives a complementary view of this dependence in terms of the separation between the two resonances: moving up through parameter space will increase $\frac{r_1}{r_2}$ up to some maximum value, so that the resonances move closer together up to some minimum separation, before moving apart again. The maximum $\frac{r_1}{r_2}$ is very close to 1 when $\frac{a_1}{a_2}\ll1$ -- i.e. the distance between the resonances approaches zero, such that one seems to pass through the other, which we can see from the fact that the contours towards the left of Fig.~\ref{fig:srloc} look like pairs of straight lines crossing each other.

Next, let us fix the \textit{mass} ratio and move from left to right in parameter space, by increasing $a_1$. From Table~\ref{tab:limits} and Fig.~\ref{fig:time-scale}, this causes both eigenfrequencies $g_i$ to increase, and hence both time-scales $\tau_i$ to decrease. In regions II and III, since $A(a)$ stays constant, both $r_1$ and $r_2$ thus move inwards. In region I, $A(a)$ increases slightly faster with $x$ than does $g_2$, so that $r_2$ in fact moves slightly outwards as $\frac{a_1}{a_2}$ approaches unity.

This complicated set of dependences can be summarised on a simplistic level as follows. The ratio $\frac{m_2}{m_1}$ primarily controls the relative separation between the two resonances -- in Fig.~\ref{fig:r1r2} this translates into the contours having quite a shallow gradient -- with more extreme mass ratios pushing the resonances apart. This is because the $r_i$ are set by the intersection of $A(a)$ with $g_i$, and the $g_i$ are only of comparable magnitude where $\frac{m_2}{m_1}$ is comparable with $\sqrt{\frac{a_1}{a_2}}$. The ratio $\frac{a_1}{a_2}$ primarily controls the absolute locations of the resonances; they are closer to the planets when the planets are closer together. This is explained by the fact that bringing the inner planet closer to the outer always increases both eigenfrequencies; since $A(a)$ is monotonically decreasing, the intersection points move closer in.

\subsection{Resonance Widths}
\label{sec:sr_widths} 

Another property that characterises the secular resonances is the range of semi-major axes over which they significantly increase planetesimal eccentricities -- that is, their widths. This is not as straightforward to quantify as the locations and time-scales, because in this theory the forced eccentricity becomes infinite at each $r_i$. One way to define the width of a resonance is to simply calculate the distance over which the forced eccentricity is above some threshold value $e_0$. Given that $e_{\mathrm{forced}}=\sqrt{k_0^2+h_0^2}$, equation~(\ref{eqn:h0k0}) can be used to show that the \textit{time-averaged} square of the forced eccentricity is

\begin{equation}\label{eqn:mean_ef}	
	\langle e_{\mathrm{forced}}^2 \rangle = \sum_{i=1}^{2}\left(\frac{\nu_i}{A-g_i}\right)^2.
\end{equation}

\begin{figure}
	\centering
    \hspace{-0.5cm}
	\includegraphics[width=0.5\textwidth]{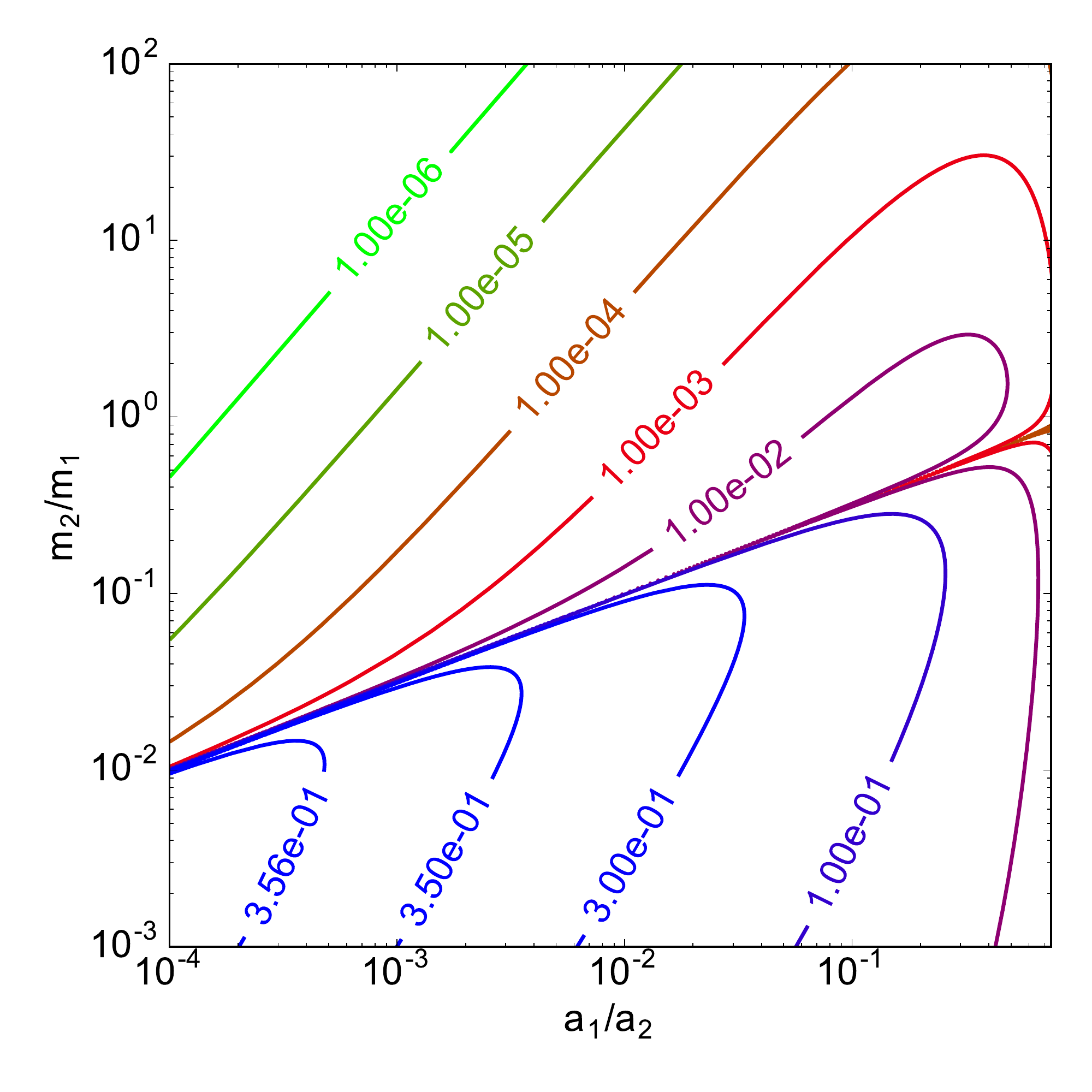}
	\caption{Contour plot showing the ratio of the width of the innermost resonance to the semi-major axis of the outermost planet, i.e. $\frac{w_1}{a_2}$, with $E_1=E_2=0.1$ and $e_0=0.2$. The innermost resonance is very narrow above the line $y=\sqrt{x}$.}
	\label{fig:widths1}
    
    \hspace{-0.5cm}
	\includegraphics[width=0.5\textwidth]{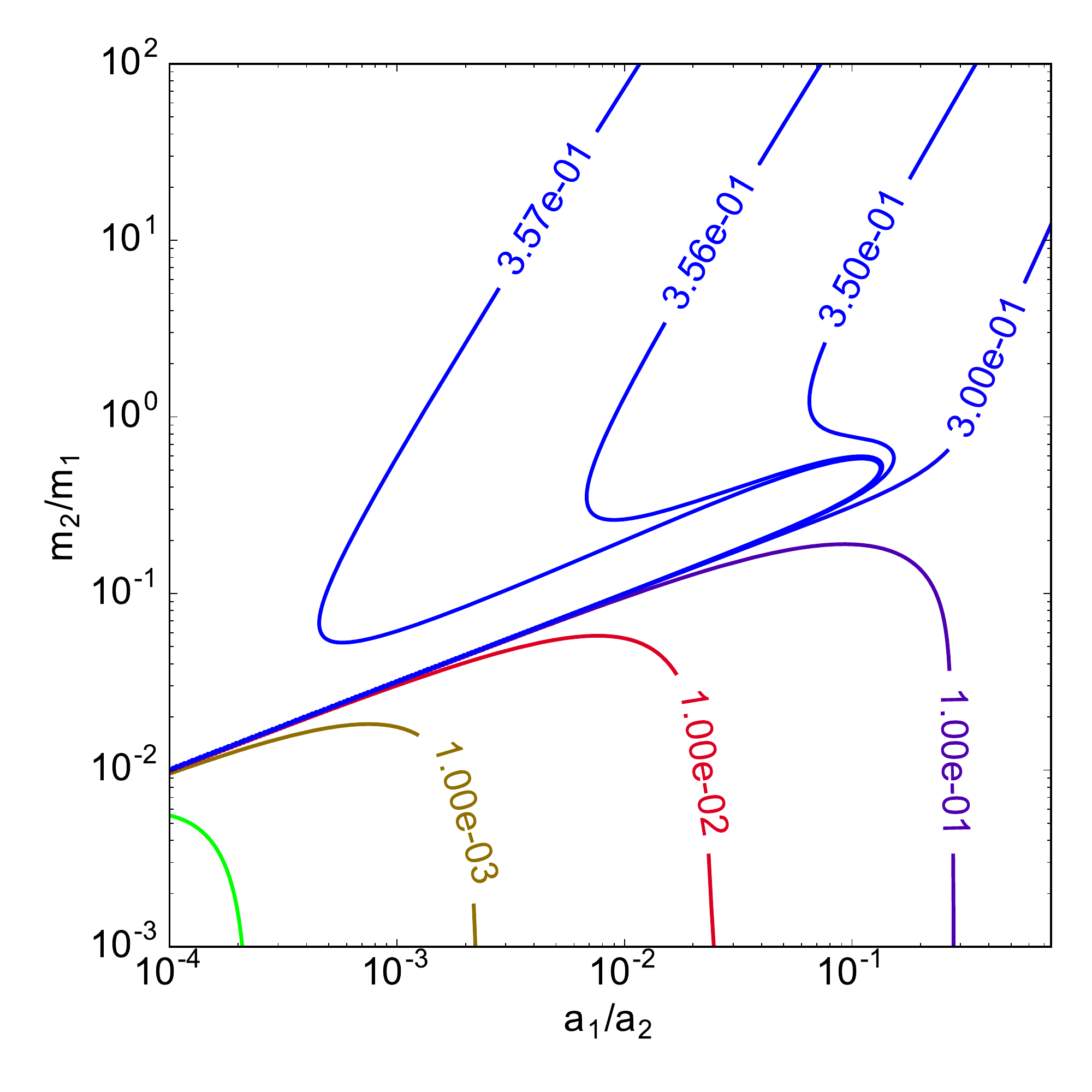}
	\caption{Contour plot showing the ratio of the width of the outermost resonance to the semi-major axis of the outermost planet, i.e. $\frac{w_2}{a_2}$, with $E_1=E_2=0.1$ and $e_0=0.2$. The contour in the lower left corner takes the value $10^{-4}$. The outermost resonance is very narrow below the line $y=\sqrt{x}$.}
	\label{fig:widths2}
\end{figure}

Near the $i^{\mathrm{th}}$ resonance, the $i^{\mathrm{th}}$ term in the sum dominates; thus, we can estimate the range of semi-major axes near $r_i$ over which $e_{\mathrm{forced}}$ exceeds $e_0$ by finding the two values of $a$ that satisfy

\begin{equation}\label{eqn:width_threshold}	
	\left(\frac{\nu_i (a) }{A(a)-g_i}\right)^2=e_0^2
\end{equation}

and taking the difference between them to obtain the width $w_i$. The result depends on the initial conditions of the system -- specifically, the initial eccentricities and longitudes of pericentre of the planets -- since these control the scaling of the eigenvectors $e_{ji}$, and therefore affect $\nu_i$ via equation~(\ref{eqn:nui}). We will assume that the longitudes of pericentre were both zero initially, and we denote the initial eccentricities by $E_j$. This leads to the following expression for $\nu_i$:

\begin{equation}\label{eqn:nui_explicit}
\nu_i=\left(A_1+\frac{A_2A_{21}}{g_i-A_{22}}\right)\left|\frac{(g_1-A_{22})(g_2-A_{22})}{A_{21}(g_1-g_2)}\left(\frac{A_{21}E_1}{g_k-A_{22}}-E_2\right)\right|,
\end{equation}

where $k\neq i$. The ratio $\nu_i (a) /(A(a)-g_i)$ depends only on $\frac{a_1}{a_2}$, $\frac{m_2}{m_1}$ and $\frac{a_2}{a}$, which means that for some fixed $E_j$, equation~(\ref{eqn:width_threshold}) can be solved to give $\frac{w_i}{a_2}$ in terms of only the ratios of planetary masses and semi-major axes. The results of doing so numerically are displayed in Figs. \ref{fig:widths1} and \ref{fig:widths2}, which correspond to a choice of $E_1=E_2=0.1$ and $e_0=0.2$. The most striking feature of these plots is that there appear to be two distinctly different regimes, separated by the line $y=\sqrt{x}$. Above this line, $\frac{w_2}{a_2}$ does not depend strongly on $\frac{a_1}{a_2}$ or $\frac{m_2}{m_1}$; below it, $\frac{w_2}{a_2}$ is much smaller than above, becoming smaller towards the left hand side of the plot. The reverse is true for $\frac{w_1}{a_2}$, which is roughly constant below the line and takes comparatively small values above it.

From Fig.~\ref{fig:r1r2}, this means that if the resonances are far apart, one of them will be much narrower than the other -- the widths can only be comparable if the resonances are close to each other. The exception to this is if the planets lie in the lower-right corner of parameter space, where the resonance at $r_2$ is the `narrow' one -- here, $r_1$ approaches $a_2$, which `squashes' the resonance at $r_1$ so that it is narrow too.

As with Figs.~\ref{fig:srloc}--\ref{fig:time-scale}, we can explain the width contours analytically by examining limits in regions I, II and III. The widths $w_i$ can be approximated from $\nu_i$ using the fact that

\begin{equation}\label{eqn:dAda_approx}
A\left(r_i+\frac{w_i}{2}\right)-A\left(r_i-\frac{w_i}{2}\right)\approx \frac{\mathrm{d}A}{\mathrm{d}a}\Bigr|_{r_i} w_i.
\end{equation}

Equation~(\ref{eqn:width_threshold}) tells us that $A(r_i\pm w_i/2)\approx g_i\mp|\nu_i(r_i)|/e_0$, and thus

\begin{equation}\label{eqn:wi_approx}
w_i\approx \frac{-2|\nu_i(r_i)|}{e_0\frac{\mathrm{d}A}{\mathrm{d}a}\Bigr|_{r_i}} \approx \frac{4|\nu_i(r_i)|}{7e_0g_i}r_i,
\end{equation}

where we used the fact that $\frac{\mathrm{d}A}{\mathrm{d}a}\Bigr|_{r_i}\approx-\frac{7g_i}{2r_i}$, which can be seen from the results in Table~\ref{tab:limits}. The first four columns of Table~\ref{tab:approx_widths} give the limiting forms of $|\nu_i(r_i)|$ and $\frac{\mathrm{d}A}{\mathrm{d}a}\Bigr|_{r_i}$ in each region of parameter space; these are used to calculate the approximate values of $\frac{w_i}{a_2}$ shown in the final two columns. Setting $\frac{w_i}{a_2}$ to a constant gives the slopes of the contours (apart from in the regions where one of the widths does not depend strongly on the planetary parameter ratios, in which case it gives only the approximate value of the contours in that region).

Equation~(\ref{eqn:wi_approx}) shows that the width $w_i$ increases linearly with $|\nu_i(r_i)|$. This makes intuitive sense because from equation~(\ref{eqn:mean_ef}), $|\nu_i(r_i)|$ can be thought of as the `strength' of the resonance at $r_i$, in that it controls the scaling of the forced eccentricity of particles in the vicinity of $r_i$. Apart from the arbitrarily chosen threshold $e_0$, the only other quantity that influences the widths is the gradient of $A(a)$ in the vicinity of the resonances, with smaller gradients giving larger widths. This is because if the $A(a)$ curve is relatively flat near $r_i$, the denominator of the $i^{\mathrm{th}}$ term in equation~(\ref{eqn:mean_ef}) remains small over a broad range of semi-major axes.

We can understand Figs.~\ref{fig:widths1} and \ref{fig:widths2} in terms of the way in which these two determining factors vary across the parameter space. For instance, in region I, the resonance strength $|\nu_1(r_1)|$ does not depend on $x$ or $y$, but the gradient $\frac{\mathrm{d}A}{\mathrm{d}a}\Bigr|_{r_1}$ becomes shallower as $y$ increases; the width $w_1$ therefore increases as we move up through parameter space. In region II, $|\nu_1(r_1)|$ and $\frac{\mathrm{d}A}{\mathrm{d}a}\Bigr|_{r_1}$ both scale in the same way with $x$ and $y$. Thus, when moving through that part of parameter space, any increase (or decrease) in the strength of the inner resonance is compensated for by the precession frequency gradient becoming steeper (or shallower), so that $w_1$ remains approximately constant. In region III, $|\nu_1(r_1)|$ decreases with increasing $y$, while $\frac{\mathrm{d}A}{\mathrm{d}a}\Bigr|_{r_1}$ depends only on $x$, so that in the uppermost part of parameter space $w_1$ decreases as we move up. Similar considerations explain the behaviour of $w_2$.

Table~\ref{tab:approx_widths} also offers some insight into how the widths depend on the initial planetary eccentricities. More eccentric planets lead to wider resonances. Above $y=\sqrt{x}$, the width of the innermost resonance $w_1$ is controlled only by $E_1$ and that of the outermost resonance $w_2$ by $E_2$; below, the dependence switches such that $E_1$ controls $w_2$ and $E_2$ controls $w_1$. This is to be expected since, as we concluded in the previous subsection, in the upper region of parameter space the resonance at $r_1$ is associated with the precession of the inner planet and that at $r_2$ with the outer, with the situation reversed in the lower region. While Table~\ref{tab:approx_widths} suggests that the widths are directly proportional to the eccentricities, this only applies when $E_j\ll1$, since Laplace-Lagrange theory is only second-order in the eccentricities. For eccentric planets, the limits presented will overestimate the actual widths.

\begin{table*}
\centering
\begin{tabular}{c|c|c|c|c|c|c}
	\hline 
	& $|\nu_1(r_1)|$ & $|\nu_2(r_2)|$ & $\frac{\mathrm{d}A}{\mathrm{d}a}\Bigr|_{r_1}$ & $\frac{\mathrm{d}A}{\mathrm{d}a}\Bigr|_{r_2}$ & $\frac{w_1}{a_2}$ & $\frac{w_2}{a_2}$ \\ 
	\hline 
	\rule{0pt}{10pt} $\mathrm{I:\ }y\ll x^2$ & $\frac{5}{4}PE_2$ & $\frac{5}{4}PE_1x^{33/14}y^{2/7}$ & $-\frac{7}{2}\frac{P}{a_2}\frac{x^2}{y}$ & $-\frac{7}{2}\frac{P}{a_2}x^{19/14}y^{2/7}$ & $\frac{5}{7}\frac{E_2}{e_0}\frac{y}{x^2}$ & $\frac{5}{7}\frac{E_1}{e_0}x$ \\ 
	\rule{0pt}{10pt}$\mathrm{II:\ }x^2\ll y\ll \sqrt{x}$ & $\frac{5}{4}PE_2\left(\frac{x^2}{y}\right)^{9/7}$ & $\frac{25}{16}PE_1x^{41/14}$ & $-\frac{7}{2}\frac{P}{a_2}\left(\frac{x^2}{y}\right)^{9/7}$ & $-\frac{7}{2}\frac{P}{a_2}x^{27/14}$ & $\frac{5}{7}\frac{E_2}{e_0}$ & $\frac{25}{28}\frac{E_1}{e_0}x$\\ 
	\rule{0pt}{10pt}$\mathrm{III:\ }y\gg \sqrt{x}$ & $\frac{25}{16}PE_1\frac{x^{24/7}}{y}$ & $\frac{5}{4}PE_2\left(\frac{x^2}{y}\right)^{9/7}$ & $-\frac{7}{2}\frac{P}{a_2}x^{27/14}$ & $-\frac{7}{2}\frac{P}{a_2}\left(\frac{x^2}{y}\right)^{9/7}$ & $\frac{25}{28}\frac{E_1}{e_0}\frac{x^{3/2}}{y}$ & $\frac{5}{7}\frac{E_2}{e_0}$\\ 
    \hline 
\end{tabular}
\caption{Approximate forms of $|\nu_i(r_i)|$, $\frac{\mathrm{d}A}{\mathrm{d}a}$ evaluated at $r_i$, and $\frac{w_i}{a_2}$ in each region of parameter space, where $x=\frac{a_1}{a_2}$, $y=\frac{m_2}{m_1}$, the $E_j$ are the initial planetary eccentricities and $e_0$ is the threshold eccentricity used to define the widths. The pre-factor $P$ is given by $\frac{3n_2m_2}{4m_{\mathrm{c}}}$.}\label{tab:approx_widths}
\end{table*}

\section{Parameter Space Constraints for HD~107146}
\label{sec:PS_constraints}

Given a debris disc with a depletion in surface density, we can hypothesise that this is the result of
eccentricity excitation by the secular resonances of an undetected planetary system. Observations
of the disc can then be used to place constraints on the possible masses and semi-major axes of the
proposed planets: they must be chosen in such a way that at least one of the secular resonances that could be affecting the disc structure is located at the depletion. The locations of the other secular resonances may or may not be important: some might act on time-scales that are too long, or have widths that are too narrow, to have had any observable effect on the disc. To illustrate how to identify the appropriate regions in the space of planetary masses and semi-major axes, in this section we apply the considerations of section~\ref{sec:SR_theory} to the disc of HD~107146, with the aim of deducing what kind of two-planet system we expect to be able to produce an HD~107146-like structure. 

We learnt in section~\ref{sec:sr_widths} that the resonance widths are controlled not only by the masses and semi-major axes of the planets, but also by their eccentricities. We might, therefore, hope to constrain the planetary eccentricities given the width of the observed gap. In reality, the situation is more complicated than this -- as we will discuss in section~\ref{sec:discussion}, the eccentricities control both the gap width and the level of asymmetry of the disc. High resolution ALMA data have indicated that HD~107146 is axisymmetric (\citealt{Marino2018_107146}), which means we cannot invoke planets of arbitrarily high eccentricity. It is also not clear how to translate the resonance widths as defined in section~\ref{sec:sr_widths} into physical gap widths, not least because they depend on the arbitrarily chosen threshold $e_0$. Given these difficulties, in this section we will concern ourselves only with whether the resonances have non-negligible width, without attempting to use the observed width as a constraint on the planetary eccentricities. In section~\ref{sec:discussion} we will use the results of our simulations of HD~107146 to calibrate the value of $e_0$, and discuss how this could be used to estimate the eccentricity of the outermost planet in the case of other discs for which asymmetry is not ruled out.

Before discussing the constraints that we can obtain from consideration of the secular resonances, we first show how it is possible to relate the mass and semi-major axis of the outermost planet by assuming that this planet is responsible for setting the location of the inner edge of the disc.

\subsection{Disc Truncation}
\label{sec:truncation}

Close to a planet, its mean motion resonances overlap; this causes nearby planetesimals to be placed on chaotic orbits and quickly ejected from the system. From \citet{Wisdom80_ResOverlap}, this will happen in the region $a_2-\Delta a<a<a_2+\Delta a$, where 

\begin{equation}\label{eqn:delta_a}	
	\Delta a \approx 1.3{\left(\frac{m_2}{m_{\mathrm{c}}+m_2}\right)}^{2/7}a_2.
\end{equation}

This mechanism could be responsible for setting the inner disc radius $r_{\mathrm{in}}$, which is around 30au for HD~107146 (\citealt{Ricci15_AlmaObs}). By requiring $a_2+\Delta a = r_{\mathrm{in}}$, we find an expression for the mass the planet must have in order to truncate the disc at $r_{\mathrm{in}}$, as a function of its semi-major axis:

\begin{dmath}\label{eqn:m2a2}	
	m_2=\frac{m_{\mathrm{c}}}{\left({\frac{1.3}{r_{\mathrm{in}}/a_2-1}}\right)^{7/2}-1}.
\end{dmath}

The value of $m_2$ given by this equation is only guaranteed to act as an upper limit -- if $m_2$ were any larger for a given $a_2$, the observed inner disc edge would be further out. However, to simplify the discussion in this section we assume that $m_2$ is in fact equal to that given by equation~(\ref{eqn:m2a2}), since this removes one degree of freedom from the problem.

\subsection{Secular Resonance Considerations}
\label{sec:PS_resonances}

From \citet{Ricci15_AlmaObs}, the depletion in the HD~107146 disc is located in the region of 70--80au, so we wish to place at least one of the secular resonances in that region. We will examine separately the cases in which each of the two resonances in turn is fixed at the depletion.

\subsubsection{Inner Resonance at the Depletion}
\label{sec:inner}

\begin{figure*}
	\centering
    \hspace{-1.0cm}
	\includegraphics[width=0.75\textwidth]{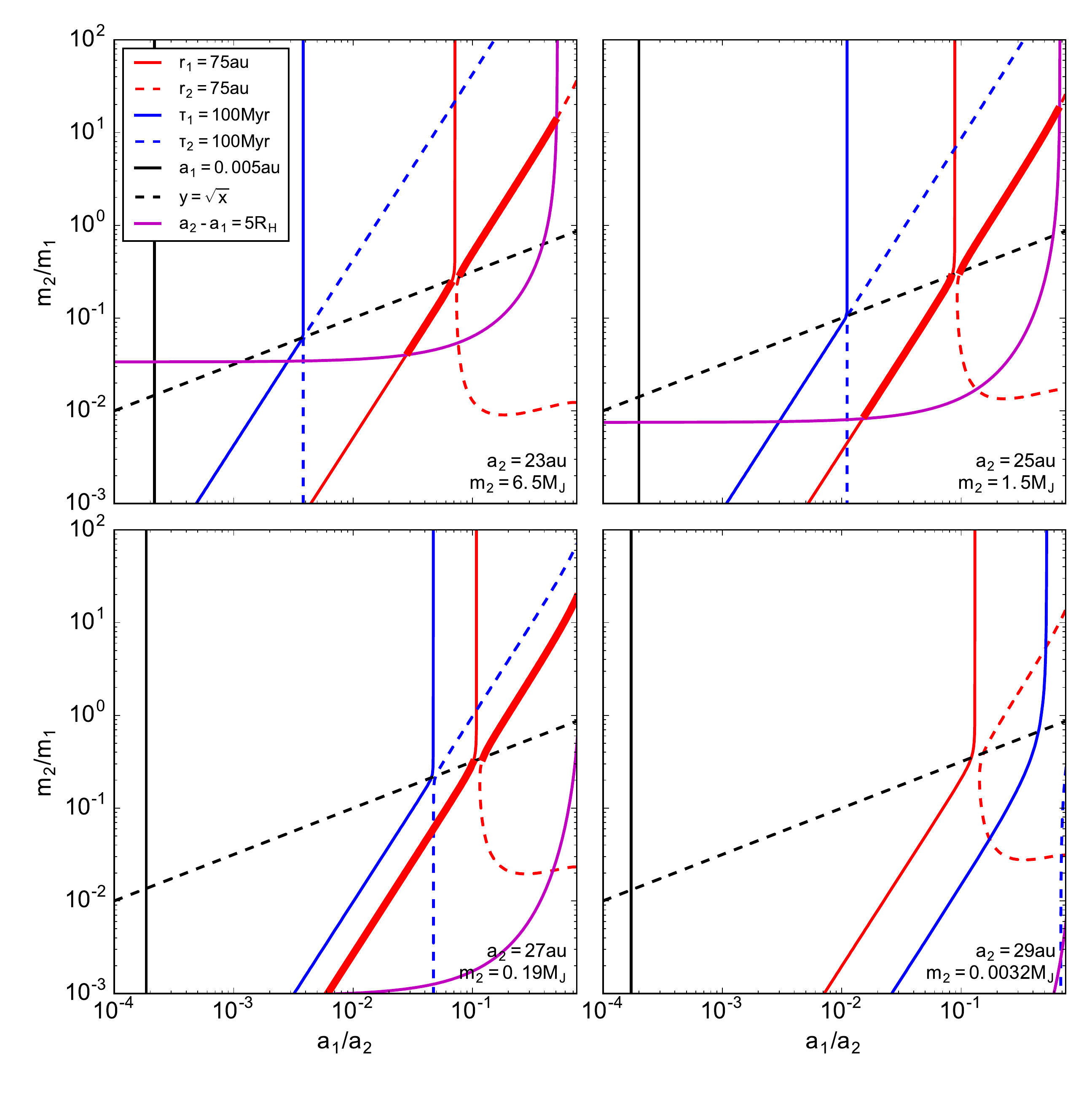}
	\caption{Graphical representation of the parameter space constraints discussed in section~\ref{sec:PS_resonances}. Each panel corresponds to the displayed values of $a_2$ and $m_2$, linked by equation~(\ref{eqn:m2a2}). The thick red lines show the loci of stable planetary parameters that place a secular resonance with non-negligible width and a time-scale less than 100Myr at 75au. There are no thick lines in the lower-right panel because the secular time-scales at 75au are too long (i.e. the blue solid and dashed lines are to the right of their respective red lines).}
	\label{fig:ps_multi}
\end{figure*}

Consider first the situation in which the \textit{inner} resonance is contributing to the depletion. We choose to fix $r_1$ at 75au; the planetary parameters must then lie along the contour $r_1=75\mathrm{au}$ in $(\frac{a_1}{a_2},\frac{m_2}{m_1})$ space. Since the quantity we are fixing is the absolute value of $r_1$ rather than its location relative to the planets (as in Fig.~\ref{fig:srloc}), one of the $a_j$ must be specified in order to plot this contour. We choose to specify $a_2$; this automatically gives $m_2$ via equation~(\ref{eqn:m2a2}). The parameter space constraints will therefore come as a series of two-dimensional $(\frac{a_1}{a_2},\frac{m_2}{m_1})$ plots, each for a given value of $a_2$; this follows because there were originally four degrees of freedom ($a_j$ and $m_j$), one of which was removed by equation~(\ref{eqn:m2a2}). The contour $r_1=75\mathrm{au}$ is displayed as a solid red line in Fig.~\ref{fig:ps_multi}, for four values of $a_2$; note that it moves to the right as we increase $a_2$. 

From \citet{Williams04_Age}, HD 107146 is $\sim$80--200Myr old. When choosing planetary parameters, it is necessary to ensure that the resulting secular time-scale is short enough that we expect the resonance to excite eccentricities significantly within the age of the star. To check approximately whether this condition is fulfilled, we draw the contour $\tau_1=100\mathrm{Myr}$ on each parameter space plot. As long as this contour lies to the left of the $r_1=75\mathrm{au}$ curve, the time-scale of the resonance should be short enough, since (from Fig.~\ref{fig:time-scale}) time-scales decrease to the right. 

The solid blue line in each panel of Fig.~\ref{fig:ps_multi} is $\tau_1=100\mathrm{Myr}$, which shifts to the right as we increase $a_2$. For $a_2\leq27\mathrm{au}$, this curve lies to the left of $r_1=75\mathrm{au}$, so we expect significant eccentricity excitation within the age of the system. However, for $a_2=29\mathrm{au}$ the outermost planet is sufficiently light (and far out) that secular effects may be too slow to have had any effect on the disc. 

We have so far made no mention of the outer resonance. One might naively expect that $r_2$ must be either just above $r_1$ (say less than 80au, so that both resonances are in the gap), or outside the outer disc edge (150au). These conditions would rule out the parts of the solid red $r_1=75\mathrm{au}$ curve where $\frac{r_1}{r_2}$ is between $\frac{75}{150}$ and $\frac{75}{80}$; the region where these conditions are satisfied could be identified by plotting contours like those in Fig.~\ref{fig:r1r2}. However, in section \ref{sec:sr_widths} we concluded that unless the resonances are very close to each other, one of them is always very narrow. In the lower region of parameter space, the narrow resonance is the one at $r_2$, and in the upper region the one at $r_1$; the dividing line between the two regimes is $\frac{m_2}{m_1}=\sqrt{\frac{a_1}{a_2}}$, the dashed black line in Fig.~\ref{fig:ps_multi}. We will therefore assume that below this line, only the inner resonance has any observable effect on the disc, and above it, only the outer. Thus, in the lower part of parameter space (where the $r_1=75\mathrm{au}$ curve is diagonal) the value of $r_2$ is irrelevant. It also follows that systems whose parameters lie on the vertical part of the $r_1=75\mathrm{au}$ contour will not in fact deplete the disc to any great degree at $r_1$, and so only the diagonal part of the contour should be considered `allowed'.

We have now identified combinations of $a_j$ and $m_j$ that configure the secular resonances appropriately -- however, some of these configurations may place the planets on unstable orbits. If the planets are too massive and/or too close to each other, one of them may be ejected from the system. We will assume that the system is unstable if the planets are within five mutual Hill radii $R_{\mathrm{H}}$ of each other, where $R_{\mathrm{H}}$ is as defined in \citet{Davies2014_Hill}:

\begin{equation}\label{eqn:hilldef}	
	R_{\mathrm{H}}=\left(\frac{m_1+m_2}{3m_{\mathrm{c}}}\right)^{1/3}\left(\frac{a_1+a_2}{2}\right).
\end{equation}

\noindent Setting $a_2-a_1=5R_{\mathrm{H}}$ gives the relation

\begin{equation}\label{eqn:hillcurve}	
	\frac{m_2}{m_1}=\left(\frac{24}{125}\frac{m_{\mathrm{c}}}{m_2}\left(\frac{1-a_1/a_2}{1+a_1/a_2}\right)^3-1\right)^{-1},
\end{equation}

which is the purple curve in Fig.~\ref{fig:ps_multi}. Parameters below and to the right of this curve are ruled out. Note that the curve depends on $m_2$, with more of the parameter space being allowed for lighter outermost planets.

Finally, the solid black line shows $a_1=0.005\mathrm{au}$; parameters should be chosen from the region to the right of this line to avoid the orbit of the inner planet crossing the star (from \citealt{Watson11_MassRad}, the stellar radius is 0.0046au). The part of the solid red $r_1=75\mathrm{au}$ curve that satisfies all of the conditions discussed in this subsection has been thickened in Fig.~\ref{fig:ps_multi}.

\subsubsection{Outer Resonance at the Depletion}
\label{sec:outer}

We could instead fix the \textit{outer} resonance at, say, $r_2=75\mathrm{au}$, and perform a similar analysis to that of section~\ref{sec:inner}. The appropriate parameters must then lie along the dashed red contour $r_2=75\mathrm{au}$ in each panel of Fig.~\ref{fig:ps_multi}. The relevant time-scale is now $\tau_2$, so secular effects should become evident within the age of the system provided the dashed red curve $r_2=75\mathrm{au}$ lies to the right of the dashed blue curve $\tau_2=100\mathrm{Myr}$; this is satisfied in all cases other than in the lower right panel ($a_2=29\mathrm{au}$). 

In the region where $\frac{m_2}{m_1}>\sqrt{\frac{a_1}{a_2}}$ the inner resonance is very narrow, which means that its location $r_1$ is irrelevant and any parameters lying along the uppermost segment of the $r_2=75\mathrm{au}$ curve are suitable (apart from those in the unstable region). Where $\frac{m_2}{m_1}<\sqrt{\frac{a_1}{a_2}}$, the outer resonance is narrow, disallowing the lower part of the $r_2=75\mathrm{au}$ curve. The allowed part of the $r_2=75\mathrm{au}$ curve has been thickened and made solid in Fig.~\ref{fig:ps_multi}.

\subsection{Summary of Constraints}
\label{sec:abs_params}

The plots of Fig.~\ref{fig:ps_multi} are useful in that they contain a lot of information about the resonances, and illustrate the origin of the `allowed' parts of parameter space. In practice, however, what we are ultimately interested in is which parts of planetary mass--semi-major axis space are expected to be able to produce an HD~107146-like depletion. Also, we chose to plot Fig.~\ref{fig:ps_multi} in terms of the \textit{ratios} of the planetary parameters since these are the more natural independent variables in the theory presented in section~\ref{sec:SR_theory} -- see for example equations (\ref{eqn:gi}) and (\ref{eqn:sr_condition}). However, for practical purposes it is perhaps more useful to plot the \textit{absolute} semi-major axes and masses of the allowed systems.

\begin{figure}
	\centering
    \hspace{-0.5cm}
	\includegraphics[width=0.5\textwidth]{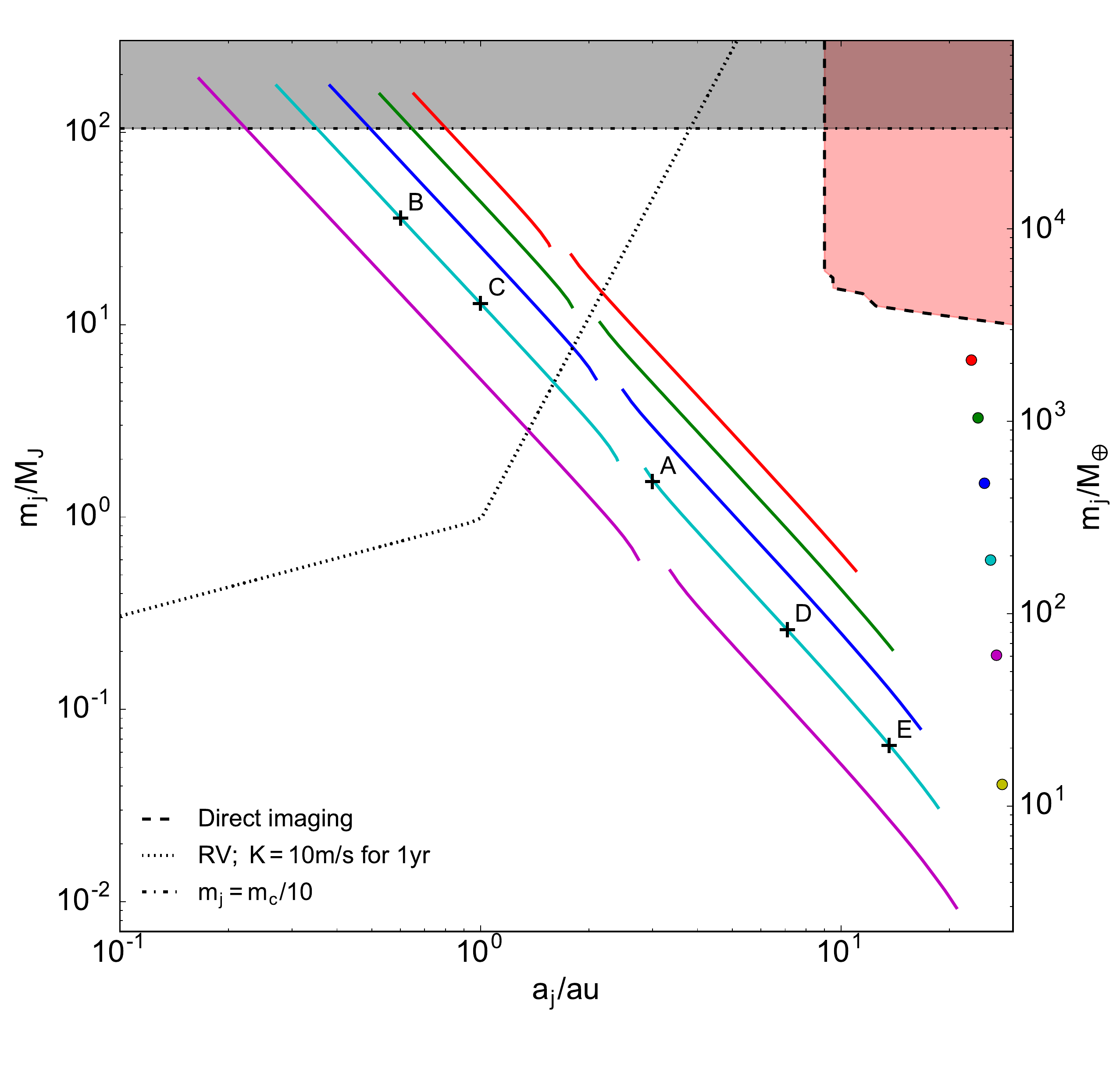}
	\caption{Plot summarising combinations of masses and semi-major axes for a two-planet system that are expected to produce a depletion at $\sim$75au. Each coloured circle is a particular choice of $a_2$ and $m_2$, linked via equation~(\ref{eqn:m2a2}). For each of these, the correspondingly coloured lines show where $a_1$ and $m_1$ can lie. The red shaded region is ruled out by direct imaging (\citealt{Apai2008_DirectImg}). The grey shaded region shows where the theoretically required companion mass exceeds one tenth of the mass of the central star. The region above the dotted line would be accessible to radial velocity measurements at a precision of 10$\mathrm{ms}^{-1}$ spanning one year. The plusses show the values of $a_1$ and $m_1$ used in the simulations discussed in section~\ref{sec:sims}.}
	\label{fig:ps_absolute}
\end{figure}

The main conclusions from the previous subsection can be summarised in a single plot in which the absolute parameters of both planets are shown, as in Fig.~\ref{fig:ps_absolute}. This plot shows a series of possible parameters for the outermost planet, chosen such that it truncates the disc at its inner edge. For each of these, the correspondingly coloured lines show the locus of possible parameters of the innermost planet. In the case of the lightest outermost planet, there are no corresponding lines because both time-scales $\tau_i$ are too long (as in the lower-right panel of Fig.~\ref{fig:ps_multi}).

Note that the allowed innermost planet parameters lie along diagonal lines in Fig.~\ref{fig:ps_absolute}. These correspond to the diagonal parts of the constant $\frac{r_i}{a_2}$ curves of Fig.~\ref{fig:srloc} (i.e. the thick red lines in Fig.~\ref{fig:ps_multi}), with the other parts of those contours giving narrow resonances. From Table~\ref{tab:approx_ratios}, therefore, the locus of allowed $a_1$ and $m_1$ for a specified $a_2$ and $m_2$ is given approximately by

\begin{equation}\label{eqn:ps_absolute_locus}	
	m_1=m_2 a_2^{11/2} r_{\mathrm{gap}}^{-7/2} a_1^{-2},
\end{equation}

where $r_{\mathrm{gap}}$ is the location where we wish to create a gap via secular resonance. The end-points of each of these lines are set by the stability limit of equation~(\ref{eqn:hillcurve}). In Fig.~\ref{fig:ps_absolute} -- in which the loci are calculated numerically rather than simply using equation~(\ref{eqn:ps_absolute_locus}) -- there is a break in each line, because the contours of constant $r_1$ and $r_2$ do not actually touch each other. 

From Table~\ref{tab:limits}, the time-scale constraint $\tau_i<t_{\mathrm{age}}$, where $t_{\mathrm{age}}$ is the age of the system, approximates to

\begin{equation}\label{eqn:ps_time-scale}	
	\left(\frac{m_2}{M_{\mathrm{J}}}\right) \left(\frac{a_2}{\mathrm{au}}\right)^2 > \frac{1}{716} \left(\frac{m_{\mathrm{c}}}{M_\odot}\right)^{1/2} \left(\frac{r_{\mathrm{gap}}}{\mathrm{au}}\right)^{7/2} \left(\frac{t_{\mathrm{age}}}{\mathrm{Myr}}\right)^{-1}.
\end{equation}

If the outermost planet parameters do not satisfy this condition, there will be no corresponding lines in Fig.~\ref{fig:ps_absolute}, i.e. there are no possible choices of innermost planet that make the resonance time-scale short enough. Equations (\ref{eqn:ps_absolute_locus}) and (\ref{eqn:ps_time-scale}) taken together could be used to make an approximate version of Fig.~\ref{fig:ps_absolute} for any other disc with an observed gap.

Fig.~\ref{fig:ps_absolute} also allows us to make links with observations. Direct imaging of HD~107146 has ruled out the shaded region in the upper right corner (\citealt{Apai2008_DirectImg}). Under the assumption that equation~(\ref{eqn:m2a2}) holds, this constrains the location of the outermost planet -- closer in than around 20au, it would enter the forbidden region. 

We are not aware of any published radial velocity limits on unseen planets for HD~107146, but it is possible to estimate the mass that such observations would be sensitive to as a function of semi-major axis. For planets with orbital periods less than the time span of the observations, the minimum detectable mass $m_{\mathrm{min}}$ as a function of semi-major axis $a$ is given by

\begin{equation}\label{eqn:rvlim}	
	\left(\frac{m_{\mathrm{min}}}{M_{\oplus}}\right)\sin{I}=11K\left(\frac{m_{\mathrm{c}}}{M_{\mathrm{\odot}}}\right)^{1/2}\left(\frac{a}{\mathrm{au}}\right)^{1/2},
\end{equation}

where $K$ is the precision of the measurements in $\mathrm{ms}^{-1}$ and $I$ is the inclination of the planet; this follows from Kepler's laws. If the time span is greater than an orbital period, the limits are less restrictive -- it is found empirically that $m_{\mathrm{min}}$ grows approximately as $a^{7/2}$ in that regime (e.g. \citealt{Kennedy2015_RVlim}). An example sensitivity curve is shown as a dotted line in Fig.~\ref{fig:ps_absolute}, corresponding to a precision of 10$\mathrm{ms}^{-1}$ and a time span of one year. This assumes an inclination $I$ of $21^{\circ}$, the same as that of the disc (\citealt{Ricci15_AlmaObs}). Some of the candidate innermost planets would be detectable using such observations, though in reality it would be difficult to precisely measure the radial velocity of HD~107146 -- as a relatively young star, it will likely have high levels of stellar activity.

The Laplace-Lagrange theory assumes that the star is much more massive than the planets, which in turn are much more massive than the planetesimals in the disc. The grey shaded area in Fig.~\ref{fig:ps_absolute} shows where $m_j>m_{\mathrm{c}}/10$, where the first of these assumptions becomes less reasonable -- thus, the results of the theory are questionable there. The theory will also be unreliable for very small $m_j$. However, this limitation is of less concern to us since the lines in Fig.~\ref{fig:ps_absolute} only extend down to a few Earth masses -- configurations involving lighter planets would have time-scales greater than the stellar age.

As discussed in section~\ref{sec:LL_theory}, the planetesimal eccentricities given by Laplace-Lagrange theory can become arbitrarily high and are thus not reliable. It is therefore necessary to perform $N$-body simulations in order to understand in detail how a disc under the influence of a system of planets `allowed' by Fig.~\ref{fig:ps_absolute} will actually look. Simulating a variety of planetary configurations will also allow us to assess how accurate the theoretical secular resonance locations are, and thus to what extent we can rely on the constraints shown in Fig.~\ref{fig:ps_absolute}. Such simulations are the subject of the following section.

\section{Simulations}
\label{sec:sims}

In section \ref{sec:PS_constraints}, we identified combinations of the semi-major axes $a_j$ and masses $m_j$ of a two-planet system that should place secular resonances in such a way that we expect to see a depletion at the same place as in observations of HD~107146. In this section, we perform numerical simulations of the dynamical and collisional evolution of the disc using some of these parameters, to ascertain whether a depleted region does indeed form at the expected location and to investigate more fully the disc structure induced by the planets.

\subsection{An Example Configuration}
\label{sec:example_sys}

To illustrate the simulation techniques, we take as an example one particular configuration. Its parameters are shown in Table~\ref{tab:exsys}. The $a_j$ and $m_j$ were chosen from Fig.~\ref{fig:ps_absolute} (in which the parameters of the innermost planet are marked with an A), and should place $r_1$ and $r_2$ at around 70 and 75au respectively; the initial eccentricities $e_j$ were chosen to be arbitrary small values.

\begin{table}
	\centering
	\begin{tabular}{c|c|c}
		$j$ & 1 & 2 \\ 
		\hline 
		$a_j$/au & 3.0 & 26.0 \\ 
		$m_j/M_{\mathrm{J}}$ & 1.5 & 0.6 \\ 
		$e_j(t=0)$ & 0.05 & 0.05 \\ 
	\end{tabular} 
	\caption{Parameters of the example planetary system considered in section~\ref{sec:example_sys}.}\label{tab:exsys}
\end{table}

\subsubsection{Dynamical Evolution}
\label{sec:dynamics}

We treat the disc as a collection of test particles moving under the gravitational influence of the star and the planets. An $N$-body integrator, REBOUND (\citealt{Rein2012_Rebound}), is used to solve for the dynamical evolution of these particles. They are started on circular ($e=0$) orbits, with semi-major axes $a$ chosen uniformly between 30 and 150au, inclinations $I$ chosen uniformly between 0 and 0.05 radians, and longitudes of ascending node $\Omega$ and true anomalies $f$ chosen uniformly between 0 and $2\pi$. The planets are started with $I_j=0$, $f_j=0$ and longitude of pericentre $\varpi_j=0$. The parameters $a_j$ and $m_j$, as well as the initial eccentricities $e_j$, are varied in further simulations described in subsection~\ref{sec:other_configs}. Simulations are run for 100Myr, with the stellar mass set to $1M_\odot$ (from \citealt{Watson11_MassRad}, the mass of HD~107146 is 1.09$M_\odot$). The integration timestep is fixed at $P_1/15$, where $P_1$ is the orbital period of the innermost planet, and we used the WHFast integration algorithm (\citealt{Rein2015_whfast}).

\begin{figure}
 	\centering
    \hspace{-0.5cm}
 	\includegraphics[width=1.\linewidth]{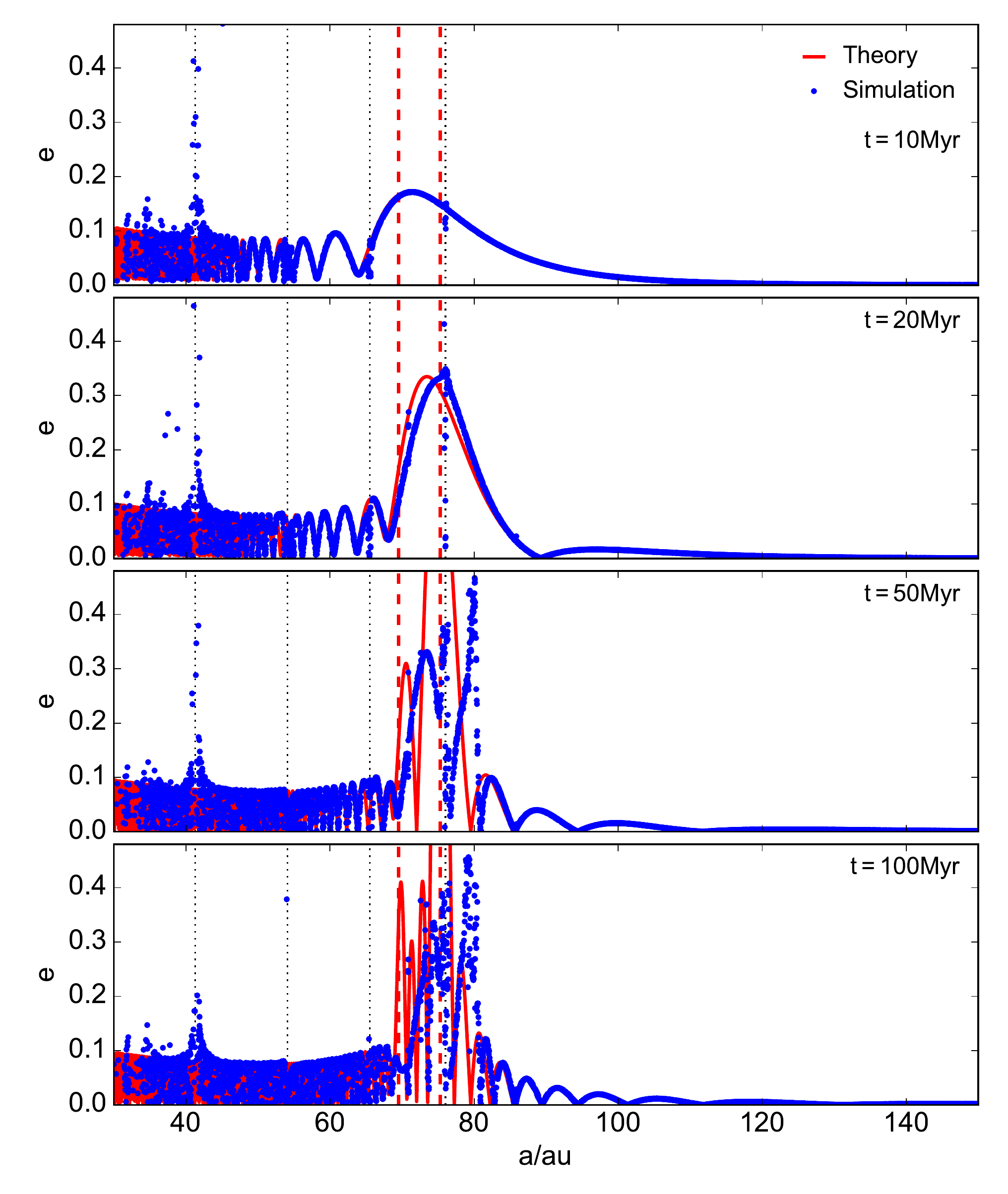}
 	\caption{Simulated and theoretical test particle eccentricity as a function of semi-major axis after 10, 20, 50 and 100Myr, for the planetary parameters shown in Table~\ref{tab:exsys}. The dotted lines show the locations of mean motion resonances with the outer planet (the effects of which are not taken into account by secular theory) and the dashed lines show the theoretical secular resonance locations. Test particle eccentricities become large in the region where they are predicted to do so, but as the Laplace-Lagrange theory is only second order in eccentricities it does not accurately reproduce the simulated profiles after many secular periods.} 
 	\label{fig:evsa_all}
 \end{figure}

We are interested in the eccentricity as a function of semi-major axis of the particles in the disc. This is shown at 10, 20, 50 and 100Myr for our example simulation in Fig.~\ref{fig:evsa_all}. We used 5000 test particles, each of which is represented by a blue dot. Also shown (as a solid line) is the eccentricity predicted by Laplace-Lagrange theory via equation~(\ref{eqn:hksol}). The locations of mean motion resonances $p:1$ with the outermost planet, where $p=2\dots5$, are shown on the plots; the secular theory is not accurate there, since it neglects the resonant terms of the disturbing function. This failure is seen most clearly at around 41au, the location of the strong $2:1$ mean motion resonance, where particles have their eccentricities increased significantly. Elsewhere, the theory and simulations are in excellent agreement after 10Myr. After many precession time-scales -- in this case, these are $\tau_1=7\mathrm{Myr}$ and $\tau_2=10\mathrm{Myr}$ -- the theory works well far from the resonances and has approximately the right eccentricity amplitude near them, but fails to reproduce the detailed structure seen in the region of $\sim$70--80au. It is unsurprising that the theory is not accurate here, since this is where the planetesimal eccentricities are highest; the Laplace-Lagrange theory is only second order in the eccentricities and predicts that they can exceed unity, violating the conservation of energy. We conclude that Laplace-Lagrange theory is useful for giving the approximate region where eccentricities become large (so that Fig.~\ref{fig:ps_absolute} is indeed useful), but since the details of the eccentricity evolution are not correct we will only compare simulations with the predicted secular resonance locations in the remainder of this paper.

\begin{figure*}
	\centering
	\includegraphics[width=0.95\textwidth]{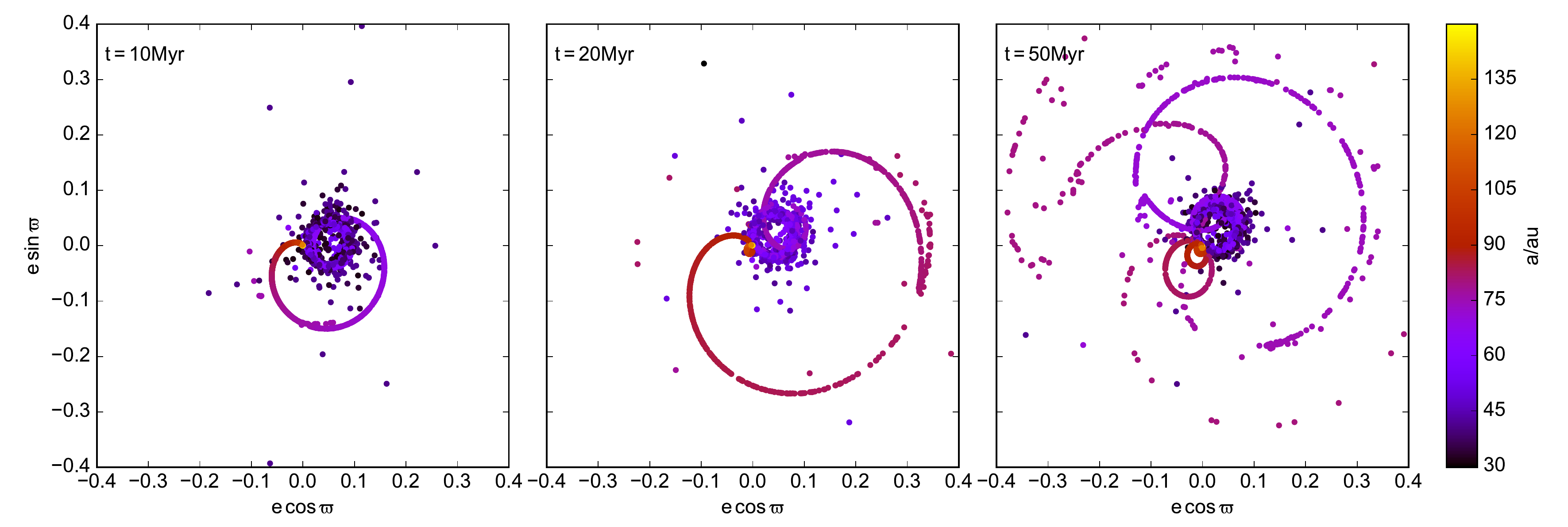}
	\caption{Simulated $(e\cos\varpi,e\sin\varpi)$ distribution of particles after 10 (left), 20 (middle) and 50Myr (right). The test particles are coloured by semi-major axis $a$. Particles near the inner disc edge, where $A(a)<g_i$, trace out a circle, while those further out, where $A(a)>g_i$, trace out a spiral. At intermediate semi-major axes where $A(a)$ is comparable to $g_i$, the particles make more complicated loop structures. These three different regimes can be understood qualitatively using the Laplace-Lagrange description (section~\ref{sec:LL_theory}) in which particles circulate around the forced eccentricity vector.}
	\label{fig:khspace}
\end{figure*}

Whilst the eccentricity profiles are of practical interest, since it is increased eccentricities that are responsible for the depletion, it is perhaps more natural to look at the $k=e\cos\varpi$ and $h=e\sin\varpi$ values of the particles, since in the Laplace-Lagrange theory these are the fundamental quantities from which $e$ is derived. Plotting these values is also instructive as they contain more information (i.e. the pericentre orientation) than the eccentricity alone. The simulation data in the $(k,h)$ plane are shown in Fig.~\ref{fig:khspace}, at 10, 20 and 50Myr. The curves that the particles lie along in Fig.~\ref{fig:khspace} can be understood in terms of the theoretical description in section~\ref{sec:LL_theory}. Towards the inner disc edge they form a circular structure centred on the forced eccentricity vector in that part of the disc, because the precession frequency $A(a)$, i.e. the rate at which particles move in circles in $(k,h)$ space, is much faster than the eigenfrequencies $g_i$, i.e. the time-scales on which the centre of the circle $(k_0,h_0)$ is moving. In the opposite limit, at the outer edge of the disc, the particles do not have time to move in complete circles before the centre moves, so the curve takes the form of a spiral. The spiral approaches the origin at large distances because the forced eccentricity is very small there. Near the secular resonances, $A(a)$ is comparable to $g_i$, which leads to more complex loop structures in the curves, because the particles are precessing around a point that is moving at the same rate as that of the precession. Because there are clear structures in $(k,h)$ space, we can expect to see non-axisymmetric structure in synthetic images generated from this simulation.

In order to examine the depletive effect of the secular resonances, we can use the simulation data to make a radial surface density profile. To do this, we split the disc into annular bins, then find the mass in each bin and divide by its area to get the azimuthally averaged surface density $\Sigma(r)$ as a function of radial distance $r$. We must assign a mass to each particle, which in general depends on its initial semi-major axis -- or equivalently, since the particles start on circular orbits, its distance $r$ from the star. Let $m(r)$ be the mass assigned to each particle initially between $r$ and $r+\mathrm{d}r$, and let $n(r)$ be the number of particles in this distance range. Then we have the initial surface density:

\begin{dmath}\label{sigma vs r}
	\Sigma(r)=\frac{m(r)n(r)}{2\pi r\mathrm{d}r}.
\end{dmath}

In our simulations, the particles are uniformly distributed in semi-major axis, so that if $N$ is the total number of test particles and $r_{\mathrm{in}}$ and $r_{\mathrm{out}}$ are the inner and outer disc radii,

\begin{dmath}\label{n vs r}
	n(r)=\frac{N\mathrm{d}r}{r_{\mathrm{out}}-r_{\mathrm{in}}}.
\end{dmath}

Combining equations (\ref{sigma vs r}) and (\ref{n vs r}) gives the mass $m(r)$ required to produce any desired initial density profile. We will assume for simplicity an initially flat profile, such that $\Sigma(r)$ is a constant, $\Sigma_0$:

\begin{dmath}\label{sigma0}
	\Sigma_0=\frac{M_{\mathrm{tot}}}{\pi(r_{\mathrm{out}}^2-r_{\mathrm{in}}^2)},
\end{dmath}

where $M_{\mathrm{tot}}$ is the initial total mass of planetesimals in the collisional cascade. This leads to an initial mass assignment proportional to $r$:

\begin{dmath}\label{m vs r}
	m(r)=\frac{2M_{\mathrm{tot}}r}{N(r_{\mathrm{out}}+r_{\mathrm{in}})}.
\end{dmath}

After dynamically evolving the system, we spread the mass of each particle around its orbit in a way that takes account of the fact that particles spend more time at apocentre than pericentre; this effectively enhances the resolution of the simulation. For each of the $N$ particles, we spawn $N_{\mathrm{sp}}$ new particles with the same $a$ and $e$; the new particles each have mass $m_{\mathrm{p}}/N_{\mathrm{sp}}$, where $m_{\mathrm{p}}$ is the mass of the parent particle, and they are given uniformly distributed mean anomalies $M$. We then numerically solve Kepler's equation,

\begin{dmath}\label{keplereq}
	M=E-e\sin E,
\end{dmath}

\noindent for the eccentric anomaly $E$ of each spawned particle. Finally, we use the standard relation

\begin{figure}
 	\centering
    \hspace{-0.5cm}
 	\includegraphics[width=1.\linewidth]{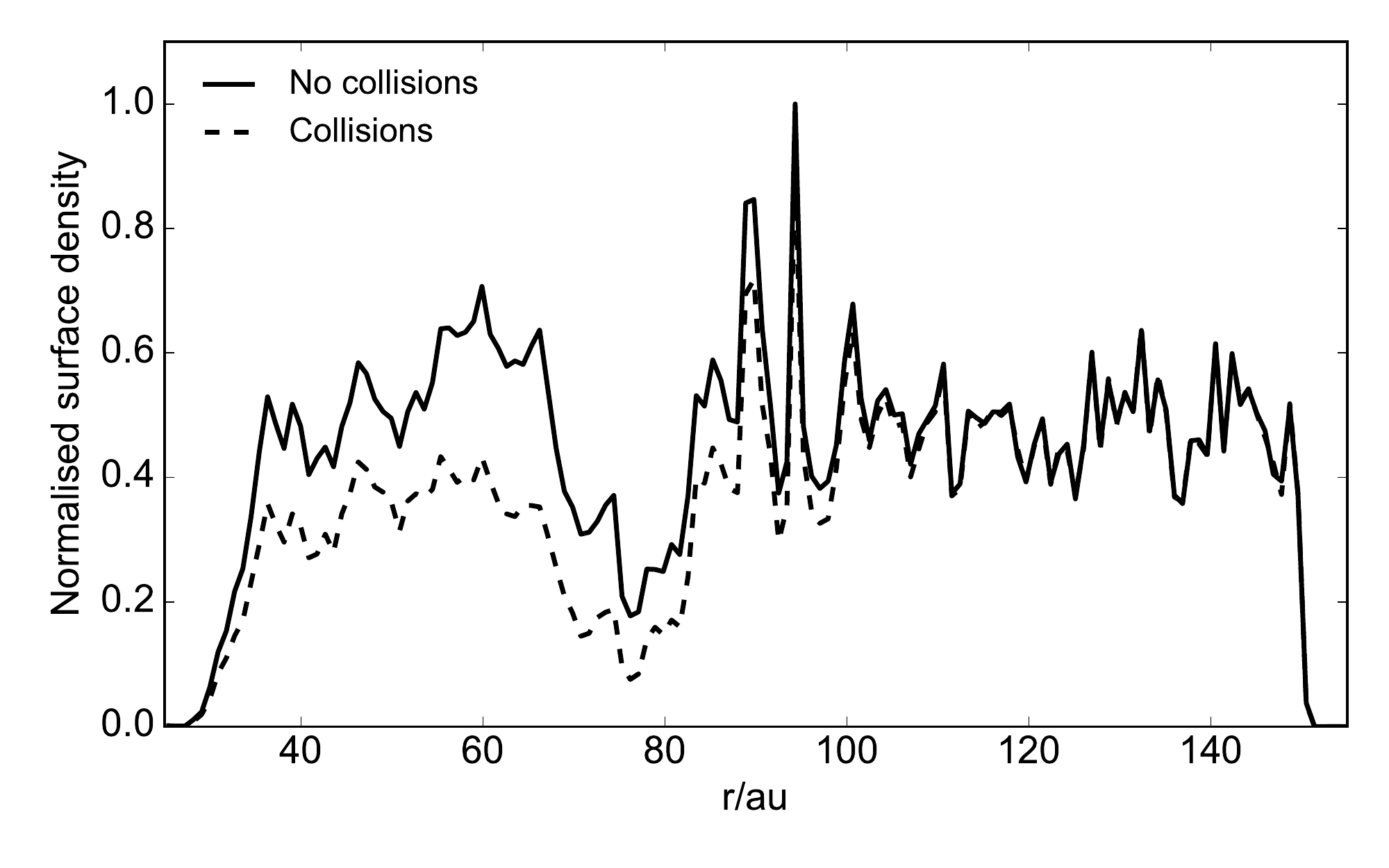}
 	\caption{Normalised surface density as a function of distance from the star after 100Myr of dynamical evolution assuming an initially flat profile, using 150 radial bins and $N_{\mathrm{sp}}=100$. The solid line includes only dynamical evolution, while the dashed line includes collisional depletion as described in section~\ref{sec:collisions}. In both cases there is a clear depletion in the vicinity of the secular resonances, which are at approximately 70 and 75au. }
 	\label{fig:density_profile}
 \end{figure}

\begin{dmath}\label{r vs E}
	r=a(1-e\cos E)
\end{dmath}

to obtain the distance of each particle from the star. Since the test particles were started with small inclinations ($I<0.05$), we treat them as coplanar. We can find $\Sigma(r)$ at the end of the disc's evolution given the resulting $N~\times~N_{\mathrm{sp}}$ particle distances and associated masses. A normalised surface density profile for our example system, assuming the profile was initially flat, is shown as a solid line in Fig.~\ref{fig:density_profile}. Because this is normalised, the absolute disc mass $M_{\mathrm{tot}}$ does not need to be known to make this plot -- however, the value of $M_{\mathrm{tot}}$ will be important when we come to include collisions in the following subsection. 

We see a clear depletion at the location where the system was designed to place one. The imprint of the initial conditions is also clear, with the overall shape of the profile being flat, though some structure is introduced by the subsidiary peaks of the eccentricity profile and by mean motion resonances. 

\begin{figure}
	\centering
    \hspace{-0.5cm}
	\includegraphics[width=1.\linewidth]{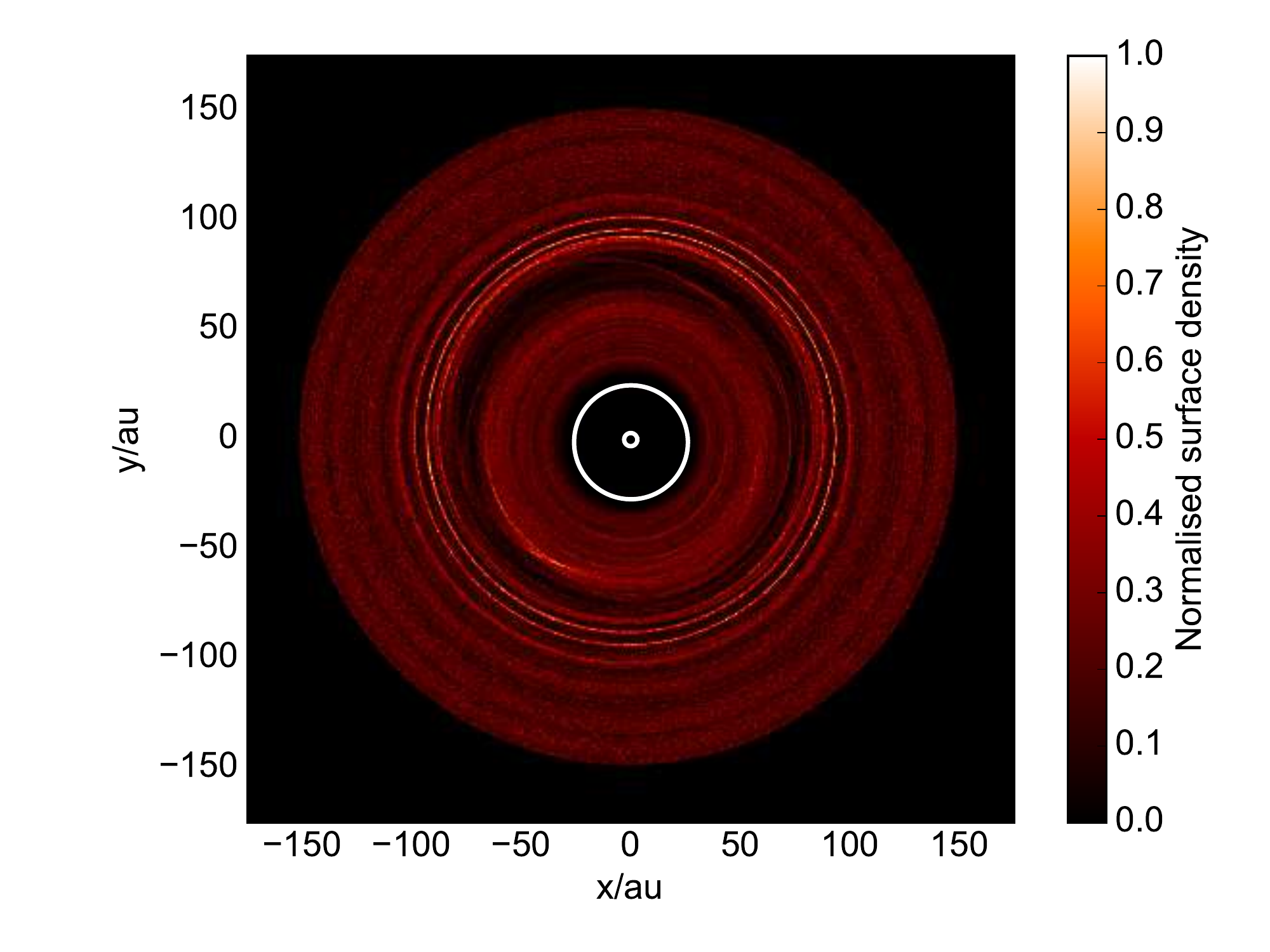}
	\caption{Image showing the normalised surface density of the disc after 100Myr of dynamical evolution assuming an initially uniform density, with $400~\times~400$ pixels and $N_{\mathrm{sp}}=300$. The white ellipses show the instantaneous planetary orbits at 100Myr. There is a depletion at around 75au due to the secular resonances; note also the spiral structure and offset inner ring. }
	\label{fig:density_img}
\end{figure}

The procedure for making Fig.~\ref{fig:density_profile} can be extended into two dimensions by binning the particles in their Cartesian coordinates in the disc plane $x$ and $y$, and spawning child particles with the same $a$, $e$ and $\varpi$ for each particle in the simulation; Fig.~\ref{fig:density_img} shows the result of doing so. This image reveals that the planets induce a spiral structure just exterior to the depletion (which showed up in Fig.~\ref{fig:density_profile} as a series of narrow peaks in the density), though on a scale smaller than the resolution of current observations. This structure can be explained by the fact that, from Fig.~\ref{fig:khspace}, $e\cos\varpi$ and $e\sin\varpi$ spiral in towards zero in the region exterior to the resonances, i.e. the longitude of pericentre decreases monotonically, wrapping round from $0$ to $2\pi$. The spiral in the disc becomes less prominent towards the outer edge because the eccentricity becomes smaller as we move further out. Note also that the disc is slightly asymmetric -- the part interior to the depletion has a small offset from the centre. This can also be understood using Fig.~\ref{fig:khspace} -- particles within around 60au lie in a circle in $(k,h)$ space centred on the forced eccentricity vector $(k_0,h_0)$, which does not vary strongly with semi-major axis in that region, giving a coherent offset to the interior part of the disc. Since the centre of this circle is moving in time, the interior part of the disc in fact precesses over time, and the extent of the offset varies. The disc of HD~107146 appears to be axisymmetric, which means that in this model the planets cannot be highly eccentric, since more eccentric planets lead to higher forced eccentricities and therefore greater asymmetries.

The surface brightness profile of HD~107146 is double-peaked, with both peaks having comparable heights (\citealt{Marino2018_107146}). Since dust further away from the star is lower in temperature, this suggests that the surface density $\Sigma$ is higher at the outer peak than the inner peak. One way to achieve such a profile would be to choose an initial profile that increases with $r$, however the protoplanetary discs from which debris discs form tend to have density profiles that \textit{decrease} with distance (e.g. \citealt{Andrews2009_PPDisks}). A perhaps more natural explanation, in which the inner part of the disc is depleted by collisions, will be explored in the following subsection.

\subsubsection{Collisional Evolution}
\label{sec:collisions}

The fact that the surface density of the HD~107146 disc appears to increase with distance from the star could be a result of its collisional evolution. As the constituent planetesimals collide with each other they break into smaller fragments, eventually grinding themselves down to the blowout size $D_{\mathrm{bl}}$, at which point they are blown out of the system by radiation pressure, resulting in a loss of mass from the disc over time. The rate of mass loss due to this process is faster where the relative velocities between the colliding objects are higher. Since the Keplerian orbital velocity is proportional to $r^{-1/2}$, planetesimals are moving faster toward inner edge of the disc, so we expect more collisional depletion there. The relative velocities between planetesimals on more eccentric orbits -- such as those near the secular resonances -- will also be higher than for near-circular orbits. Including the effect of collisions should therefore not only shape the profile into one that increases with $r$, but also provide an additional mechanism for reducing the density near the resonances on top of the purely dynamical depletion we observed in section~\ref{sec:dynamics}.

The amount of collisional depletion depends on properties of the disc that are not well known -- its mass, the strength of its constituent planetesimals, the distribution of sizes of the planetesimals and the maximum planetesimal size. We will assume an equilibrium size distribution (\citealt{Dohnanyi68_CCindex}), such that

\begin{dmath}\label{eqn:sizedist}
	n(D)\propto D^{-3.5},
\end{dmath}

where $n(D)\mathrm{d}D$ is the number of planetesimals with sizes between $D$ and $D+\mathrm{d}D$. This is in conflict with \citet{Ricci15_AlmaObs}, who derived a size distribution exponent of $-3.25\pm0.09$ for $\sim$mm-sized grains. However, as they acknowledged, extrapolating such a shallow size distribution to large planetesimal sizes leads to unreasonably large disc masses, so it is likely that larger bodies follow a steeper distribution. For simplicity, we assume that equation~(\ref{eqn:sizedist}) applies for all sizes from $D_{\mathrm{bl}}$ up to the maximum size $D_{\mathrm{c}}$; then, equation~(15) of \citet{Wyatt08_Review} can be used to relate the fractional luminosity $f$ of the disc (i.e. the ratio of the disc luminosity to the stellar luminosity) to the mass $M_{\mathrm{tot}}$. Using the value $f=1.2\times 10^{-3}$ from \citet{Williams04_Age} gives the relation

\begin{dmath}\label{eqn:Mtot_vs_Dc}
	\frac{M_{\mathrm{tot}}}{M_\oplus}=34.3\left(\frac{D_{\mathrm{c}}}{\mathrm{km}}\right)^{1/2},
\end{dmath}

where we took the radius of the disc to be its midpoint 90au, and used $D_{\mathrm{bl}}=1.7\mu\mathrm{m}$ from \citet{Ricci15_AlmaObs}. This relation removes one degree of freedom from the problem.

We adopt the model of \citet{Wyatt10_Collisions}, which also assumes an equilibrium single power law size distribution, to estimate how much mass is lost by each particle over time due to collisions. This is a post-processing of the $N$-body output, involving binning the particles in pericentre distance $q=a(1-e)$ and apocentre distance $Q=a(1+e)$, calculating the collision rate of each particle with particles in each of the $(q,\,Q)$ cells using equation~(33) of \citet{Wyatt10_Collisions}, and depleting the mass of each particle at each time according to their equation~(34). Note that by binning particles in $q$ and $Q$ without any consideration of orientation, this model assumes that the disc is axisymmetric -- which we know from section \ref{sec:dynamics} is not strictly the case here -- and thus ignores any azimuthal dependence of the collision rate. We use the shorter of $\tau_1$ and $1/R_{\mathrm{cc}}^{\mathrm{max}}$ as the timestep, where $R_{\mathrm{cc}}^{\mathrm{max}}$ is the instantaneous catastrophic collision rate of the fastest evolving particle.

\begin{figure}
	\centering
    \hspace{-0.5cm}
	\includegraphics[width=1.\linewidth]{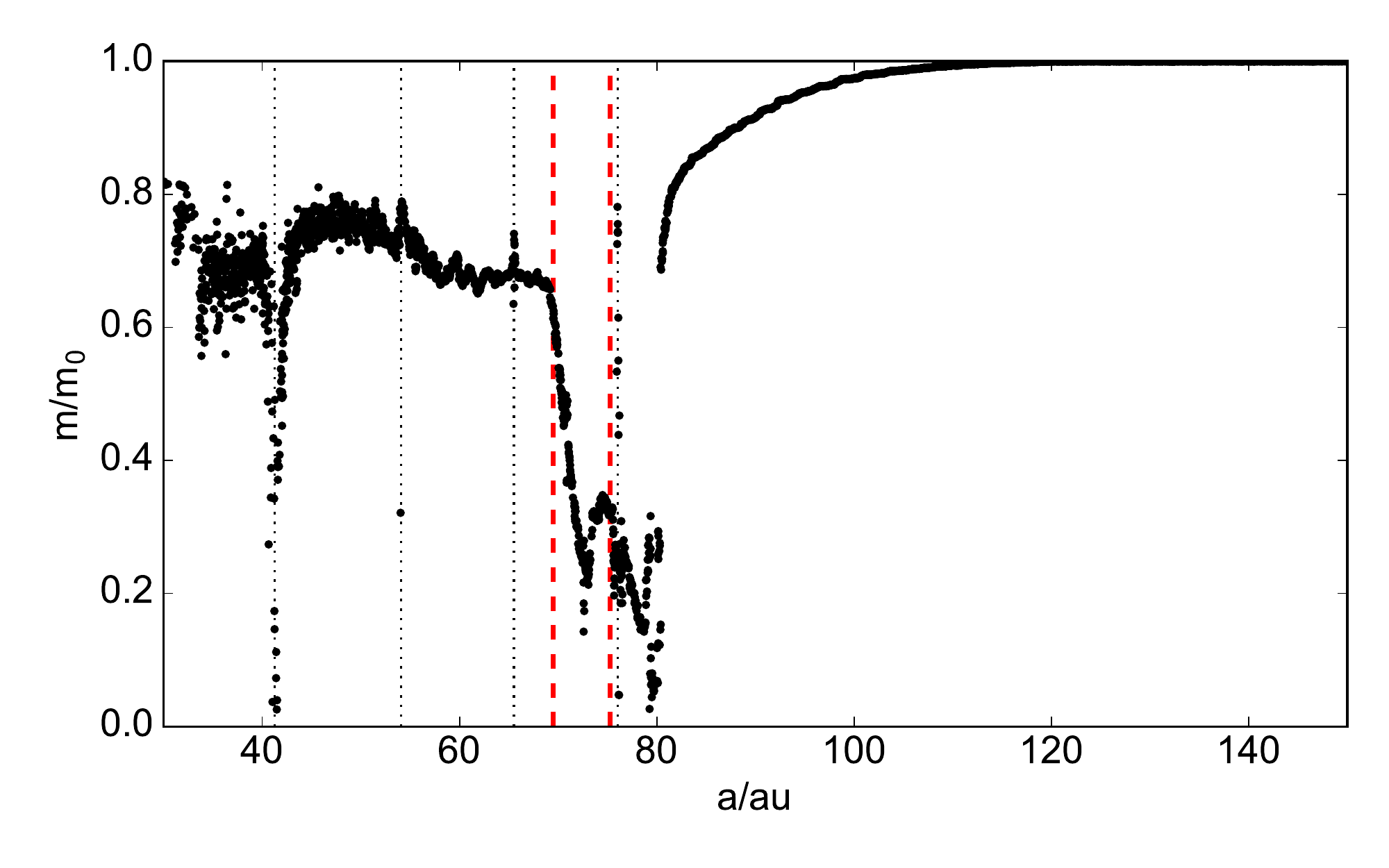}
	\caption{Fraction of initial particle mass remaining after 100Myr as a function of semi-major axis, assuming an initially flat density profile, with $Q_{\mathrm{D}}^{\star}=250\mathrm{J\,kg}^{-1}$, $D_{\mathrm{c}}=60\mathrm{km}$ and $\rho=2700\mathrm{kg\,m}^{-3}$. The dotted lines show mean motion resonance locations and the dashed lines show the secular resonance locations. The inner ring of the disc is depleted in mass relative to the outer, where the collisional lifetime is longer. Particles near the secular resonances are depleted as their enhanced eccentricities shorten their collisional lifetime. }
	\label{fig:mass_factor}
\end{figure}

Fig.~\ref{fig:mass_factor} shows the fraction of initial mass remaining after 100Myr as a function of semi-major axis, assuming an initially flat density profile, a planetesimal density $\rho$ of $2700\mathrm{kg\,m}^{-3}$, a planetesimal strength $Q_{\mathrm{D}}^\star$ of $250\mathrm{J\,kg}^{-1}$ and a $D_{\mathrm{c}}$ of $60\mathrm{km}$ (which implies $M_{\mathrm{tot}}=265M_\oplus$). We used a grid of $100~\times~100$ logarithmically spaced cells in $(q,\,Q)$ space, with $q$ and $Q$ running from 15 to 200au. Beyond around 110au, the collisional lifetime is sufficiently long that the disc is not depleted there within the age of the system. The level of depletion of the inner part of the disc can be controlled by varying the disc parameters; the reason for choosing the parameters used to make Fig.~\ref{fig:mass_factor} in particular will be explained in the following subsection. In addition to the clear collisional depletion at the secular resonances, there are also peaks and troughs in the fractional remaining mass near mean motion resonances -- though as we will see in the following subsection, even the $2:1$ resonance (which is strongly depleted in Fig.~\ref{fig:mass_factor}) is too narrow to have any effect on observations of the disc. The surface density profile that results from including collisions is shown as a dashed line in Fig.~\ref{fig:density_profile}.

\subsubsection{Comparison with Observations}
\label{sec:profiles}

\begin{figure}
	\centering
    \hspace{-0.5cm}
	\includegraphics[width=1.\linewidth]{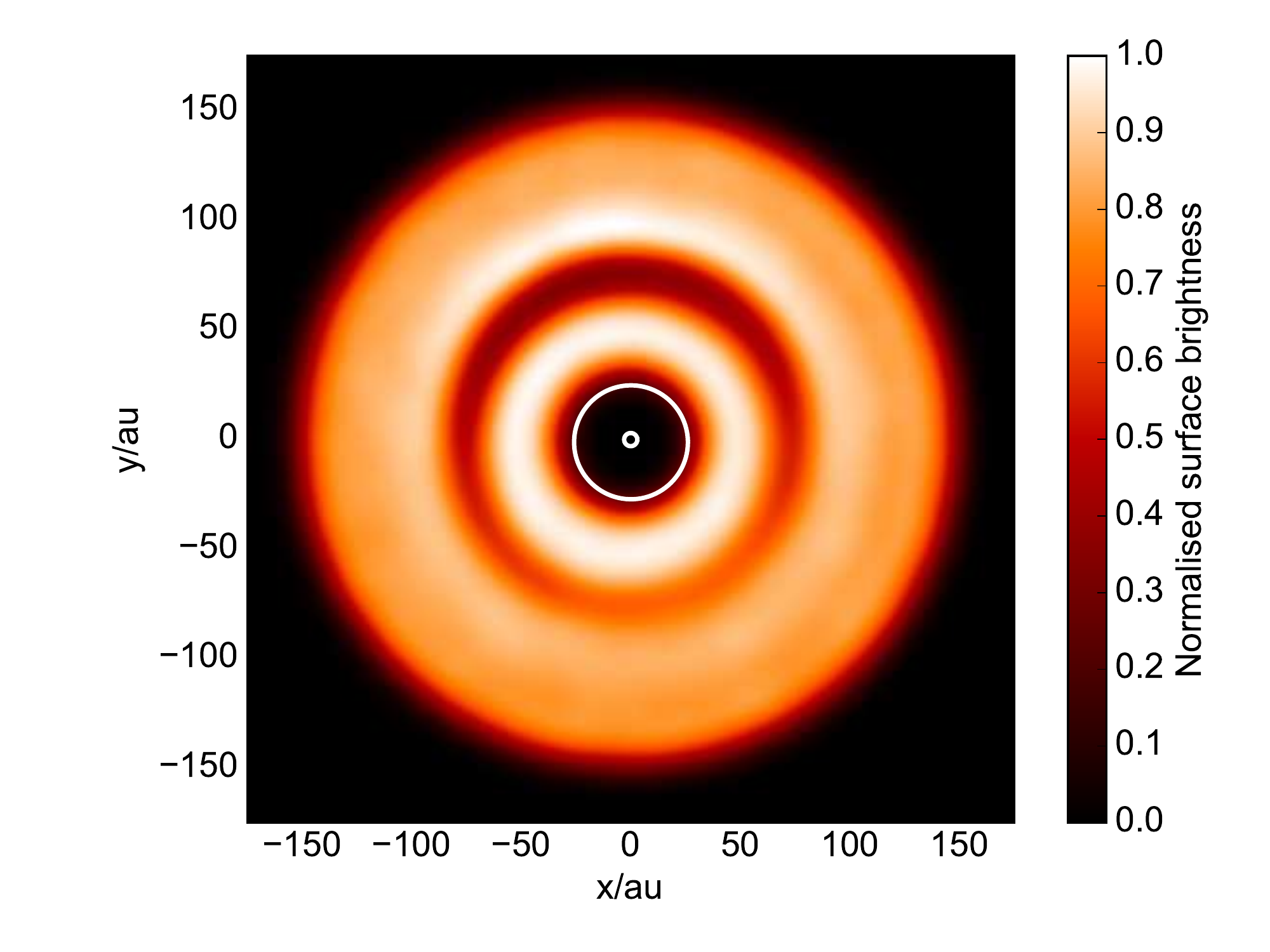}
	\caption{Synthetic image of the disc after 100Myr of dynamical and collisional evolution. The contribution from each particle in the simulation is weighted by the factor shown in Fig.~\ref{fig:mass_factor} to include the depletive effect of collisions and by $r^{-1/2}$ to convert to surface brightness. The image is convolved with a Gaussian of FWHM 18au to allow comparison with the ALMA observations of \citet{Marino2018_107146}.}
	\label{fig:disc_img}
\end{figure}

Fig.~\ref{fig:density_img} is instructive in that it gives the detailed structure of the disc, but to see what would in reality be observed, the simulation data should be convolved with an appropriate point spread function, which in the case of ALMA approximates the beam that results from CLEAN imaging of observations with some antenna configuration. For comparison with the ALMA observations of \citet{Marino2018_107146}, we convolve with a Gaussian of full width at half maximum (FWHM) 18au, the resolution of their Briggs-weighted band 6 observations. To transform surface density into surface brightness (the observed quantity), we assume that the dust grains act as black body radiators such that the contribution of each grain to the surface brightness is proportional to the black body function $B_{\nu}(T)$. Assuming they are in equilibrium with the stellar radiation, they have a temperature $T\propto r^{-1/2}$ at a distance $r$, which leads to the approximation $B_{\nu}(T)\propto r^{-1/2}$. A synthetic image is presented in Fig.~\ref{fig:disc_img}; the procedure for making this is the same as for Fig.~\ref{fig:density_img}, but the contribution from each particle is weighted by $r^{-1/2}$ as well as by the mass factor shown in Fig.~\ref{fig:mass_factor} to include the effect of collisions.

\begin{figure}
	\centering
    \hspace{-0.5cm}
	\includegraphics[width=1.\linewidth]{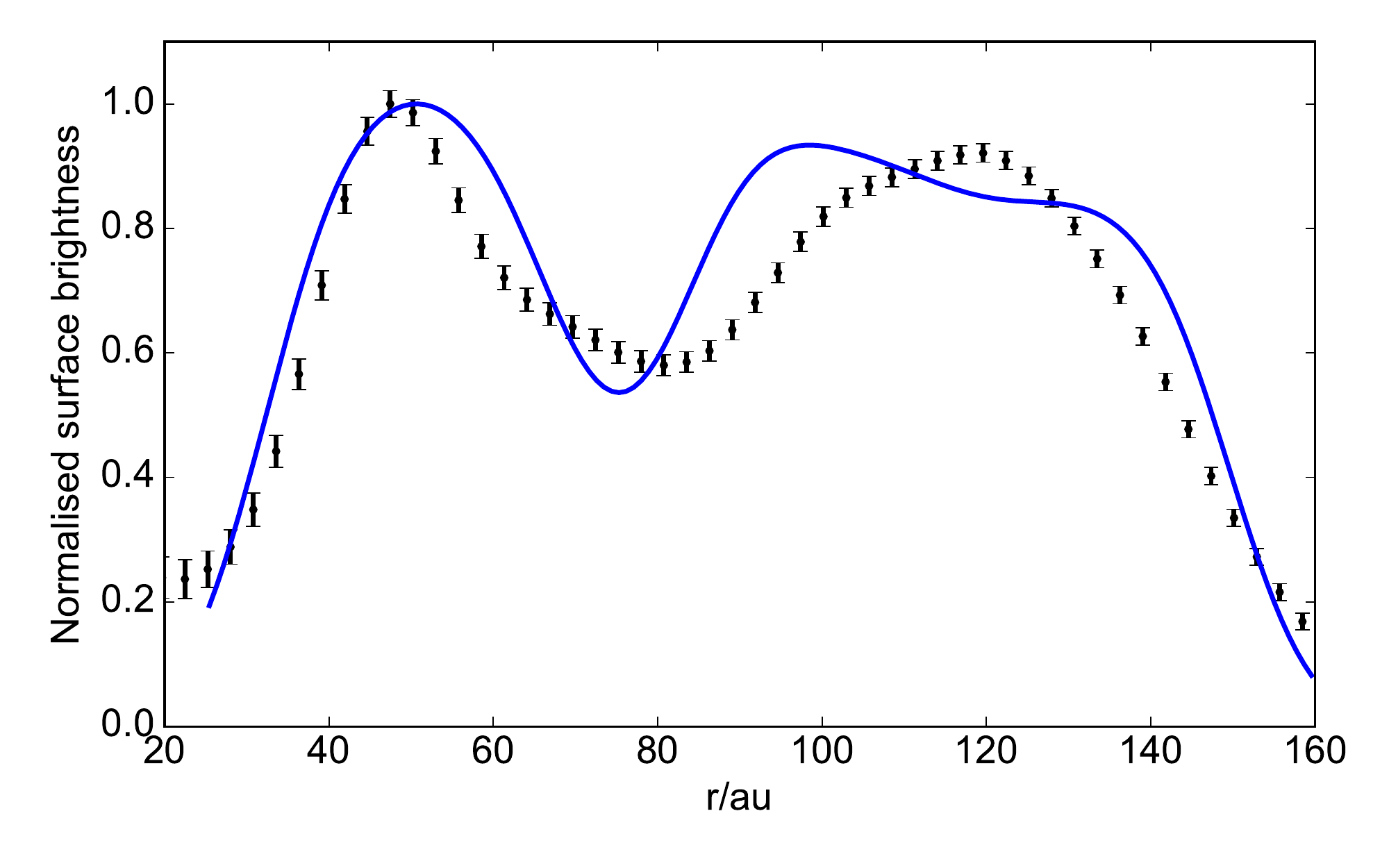}
	\caption{Azimuthally averaged radial surface brightness profile of the disc shown in Fig.~\ref{fig:disc_img} (solid line), compared with the ALMA observations (points) of \citet{Marino2018_107146}. The model profile has a depletion in the correct location and the relative peak heights are comparable with the data, though the observed gap is wider.}
	\label{fig:intensity_profile}
\end{figure}

For a quantitative comparison with data, we show the azimuthally averaged radial profile of Fig.~\ref{fig:disc_img} in Fig.~\ref{fig:intensity_profile}. Though the model is not within the observational uncertainties, this is hardly expected given that the only observational inputs to the model were the extent of the disc and the approximate location of the gap (as well as the relative brightnesses of the two rings of the disc, which, as will be explained below, were used to tune the collision rate). We do see, however, that it is possible to reproduce a disc with qualitatively the same features as in the observations -- i.e. a profile with an inner peak and a broader outer peak in the correct locations separated by a deep depletion -- just by choosing parameters from Fig.~\ref{fig:ps_absolute}. The depletion in the secular resonance model is, however, narrower than in the observed disc. We consider a way in which a wider gap could form by forcing one of the planets to migrate in section~\ref{sec:discussion}.

Note that the spiral structures seen in Figs.~\ref{fig:density_profile} and \ref{fig:density_img} are not visible at the resolution of these observations; Figs.~\ref{fig:disc_img} and \ref{fig:intensity_profile} do however make it clear that there is an excess of emission just exterior to the depletion, where the eccentric particles with semi-major axes in the depleted region will spend most of their time (i.e. near apocentre).

At this point, we can explain why the $Q_{\mathrm{D}}^\star$ and $D_{\mathrm{c}}$ chosen in the previous subsection work well. If we make a plot analogous to Fig.~\ref{fig:intensity_profile} but without including any collisional depletion, the inner peak of the brightness profile is much higher relative to the outer than in the data, with the ratio of the peak heights being around $10:6$. In the data, the relative heights are around $10:9$. Thus, if we are invoking collisions to explain the peak heights, $Q_{\mathrm{D}}^\star$ and $D_{\mathrm{c}}$ should be chosen such that the inner edge of the disc has reduced in mass by a factor of around $2/3$. To estimate some suitable parameters, we use equation~(17) of \citet{Wyatt08_Review}, which gives the time evolution of the mass of a planetesimal belt; assuming collisions became destructive early in the evolution of the disc, this equation tells us that the mass will fall to $2/3$ of its initial value after time $t_{\mathrm{c}}/2$, where $t_{\mathrm{c}}$ is the collisional time-scale. Thus, assuming an age of 100Myr, we require $t_{\mathrm{c}}$ at the inner edge of the disc to be around 200Myr. To relate $t_{\mathrm{c}}$ to $Q_{\mathrm{D}}^\star$ and $D_{\mathrm{c}}$, we use equation~(16) of \citet{Wyatt08_Review}; taking an annulus of the disc extending from 30 to 40au and setting its collisional time-scale to 200Myr then gives 

\begin{dmath}\label{eqn:QDstar_Dc}
	\left(\frac{Q_{\mathrm{D}}^\star}{\mathrm{J\,kg}^{-1}}\right)^{5/6} \left(\frac{D_{\mathrm{c}}}{\mathrm{km}}\right)^{1/2} \approx 770,
\end{dmath}

where we took the mean eccentricity in the annulus to be 0.05, consistent with Fig.~\ref{fig:evsa_all}, and assumed that the surface density profile of the disc was initially flat to calculate the mass of the annulus. The parameters used to make Fig.~\ref{fig:mass_factor} satisfy this relation, and do indeed lead to relative peak heights comparable with the data. However, note that the problem is degenerate in that we could, for example, have chosen a smaller $D_{\mathrm{c}}$ and a larger $Q_{\mathrm{D}}^\star$ and kept the same collision rate, since smaller planetesimals have shorter collisional lifetimes, which can be compensated for by making them stronger. The amount of collisional depletion required also depends on the choice of initial density profile -- for example, if one that increases with $r$ were chosen, the collisional evolution would need to be slower in order to match the observations. To summarise, we do not know the true maximum size and strength of the planetesimals, but adopting some reasonable values can deplete a disc starting with reasonable initial conditions to a level comparable with the observations.

\subsection{Other Configurations}
\label{sec:other_configs}

\begin{figure*}
	\centering
	\includegraphics[width=0.95\textwidth]{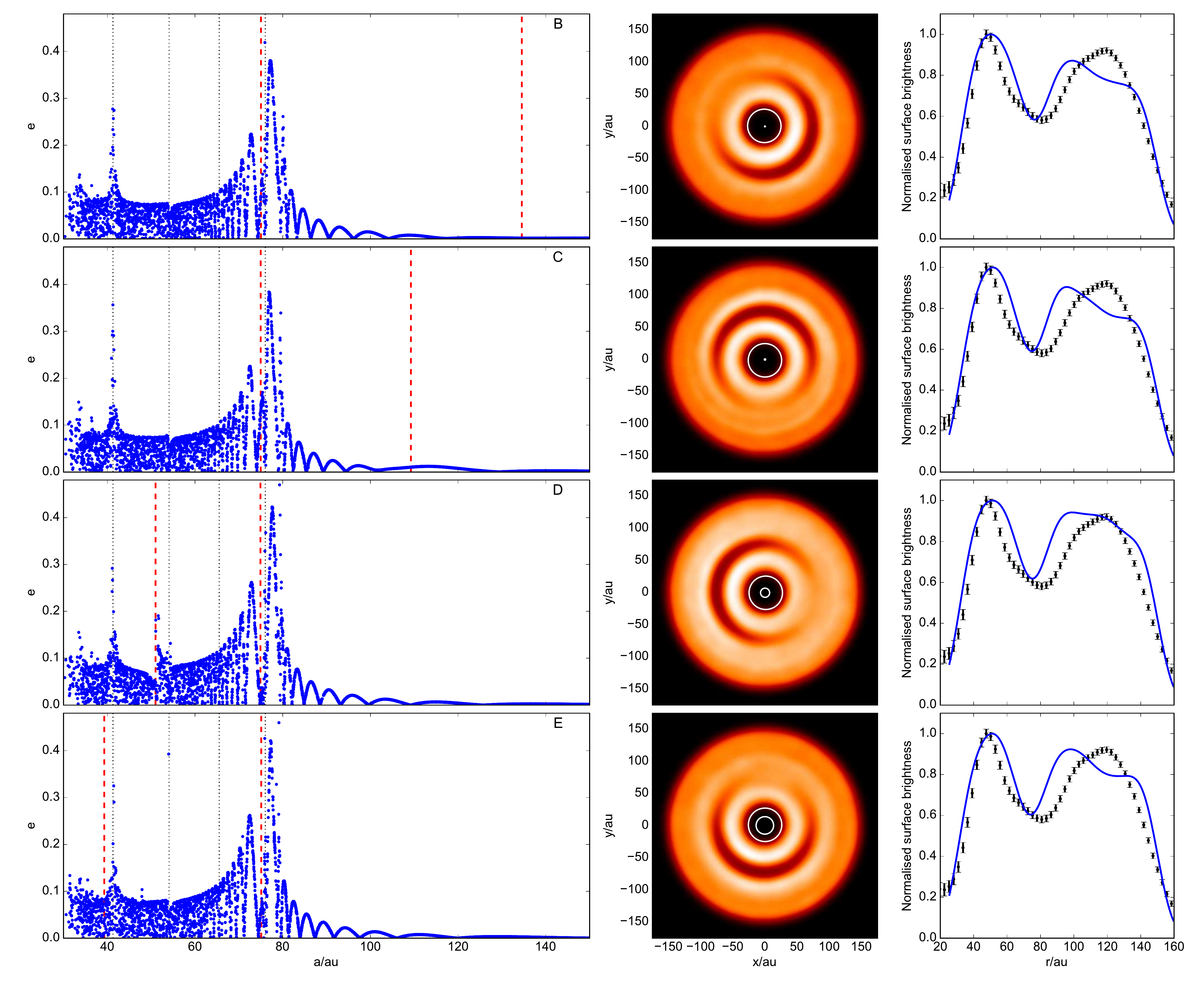}
	\caption{Summary of results of simulations B--E (as labelled in Fig.~\ref{fig:ps_absolute}), with each row corresponding to one simulation. The left hand column shows eccentricity profiles, the middle column shows the resulting synthetic images (including collisional evolution), and the right hand column shows the corresponding brightness profiles and comparison with the ALMA data of \citet{Marino2018_107146}, all after 100Myr. All simulations give similar results. }
	\label{fig:multi_sims}
\end{figure*}

We ran simulations of five planetary configurations to test whether there is a significant difference between the disc structures that result from the different parts of parameter space that are `allowed' by Fig.~\ref{fig:ps_absolute}. In all of these simulations, $a_2$ was fixed at 26au and $m_2$ at 0.6$M_{\mathrm{J}}$, such that the outer planet sets the inner disc edge. The initial planetary eccentricities were set to 0.05. Five different values of $a_1$ and $m_1$ spanning the locus of allowed parameters corresponding to our choice of outer planet were used; these are marked on Fig.~\ref{fig:ps_absolute} and labelled A--E. Simulation A is the configuration that was discussed in detail in  section~\ref{sec:example_sys}, and the results of simulations B--E are summarised in Fig.~\ref{fig:multi_sims}. All of the synthetic images and profiles shown in Fig.~\ref{fig:multi_sims} include the effects of collisional evolution with the same planetesimal density, strength and maximum size used to make Fig.~\ref{fig:disc_img}.

In each of the simulations B--E, one of the theoretical secular resonance locations is at the depletion while the other is embedded somewhere else in the disc, which distinguishes these simulations from simulation A, where both resonances were at the depletion. In the case of simulation D, the resonance at $r_2$ is producing the depletion; some eccentricity excitation is also seen at $r_1$, however, the resonance at $r_1$ is so narrow that it has no observable effect on the structure of the disc as seen by ALMA. This confirms our expectations from the results of section~\ref{sec:sr_widths} and justifies our choice to ignore the narrower resonance for the purposes of making Fig.~\ref{fig:ps_absolute}. In simulation E the inner resonance is even narrower, and there is no discernible eccentricity excitation there. For simulations B and C the resonance at $r_2$ is the narrow one; in both cases not enough time has elapsed for any effect to be seen in the eccentricity profiles there.

Even though the simulations span more than one order of magnitude in semi-major axis and two in mass, the resulting disc structures are very similar, which means that trying to identify a unique set of best-fitting parameters would not be a realistic aim. It does however validate the theory behind Fig.~\ref{fig:ps_absolute}, and shows that we will not have the problem of fine-tuning the planetary parameters to obtain the best match with the data.

\section{Discussion}
\label{sec:discussion}

In this section, we first address some issues that the secular resonance depletion model has with matching observations of HD~107146, before considering some other discs to which our model may be relevant.

\subsection{Issues for HD~107146}
\label{sec:model_issues}

\subsubsection{Asymmetry and Gap Width}
\label{sec:asymmetry}

HD~107146 appears to be axisymmetric (\citealt{Ricci15_AlmaObs}; \citealt{Marino2018_107146}), but the synthetic images in Figs. \ref{fig:disc_img} and \ref{fig:multi_sims} all show gaps that vary in width and depth with azimuthal angle, because the inner ring is offset relative to the outer. As explained in section~\ref{sec:dynamics}, this is a result of the forced eccentricity of the inner ring. This forced eccentricity is set by the planetary eccentricities; we might thus hope to reduce the asymmetry by starting the planets with lower eccentricities.

To test this, we ran a simulation with the same values of $a_j$ and $m_j$ as in simulation A (Table \ref{tab:exsys}) but with the initial eccentricities both set to 0.01. This did not, in fact, give a more HD~107146-like disc, for two reasons. Firstly, the gap became even narrower -- as expected from section~\ref{sec:sr_widths} -- which is bad from the point of view of matching radial profiles because the gap was already narrower than in the observations in Fig.~\ref{fig:intensity_profile}. Secondly -- and relatedly -- although the offset became smaller, the gap did not look any less asymmetric. This is because for a narrower gap, less of an offset is needed for the inner ring to cover the same fraction of the gap. We thus conclude that the secular resonance model is best suited to discs that show asymmetric gaps.

We also found in section~\ref{sec:sims} that while our model gives a double-ringed disc structure broadly similar to HD~107146, the depleted regions in the simulations we presented are narrower than in the observations of \citet{Marino2018_107146}. We know from section~\ref{sec:sr_widths} that making the planets more eccentric would give a wider gap. However, this would also increase the forced eccentricity in the disc, furthering the departure from axisymmetry.

One might hope to widen the depletion in the model without any increase in asymmetry by placing the resonance locations $r_1$ and $r_2$ fairly close to each other but somewhat more widely separated than in simulation A (Table~\ref{tab:exsys}). However, from the results of section~\ref{sec:sr_widths}, one of them will quickly become unimportant as they move apart, so that this is not an effective way to increase the range of distances over which eccentricities are high.

The non-axisymmetry of the secular resonance model need not always be problematic. It is, of course, possible that future observations will reveal other discs with both gaps and evidence for asymmetry. There would also be more flexibility to make wider gaps using more eccentric planets if the model were applied to a disc with lower resolution observations, which would not necessarily be able to rule out asymmetry. In such cases, it would be possible to obtain an estimate of the eccentricity required by the outermost planet by using the theory of section~\ref{sec:sr_widths}, informed by the simulations of section~\ref{sec:sims}.

The definition of resonance width used in section~\ref{sec:sr_widths} was sufficient to predict that unless the two secular resonances are very close together, one of them will be negligibly narrow -- a fact that was verified by the simulations of section~\ref{sec:sims} -- because that prediction only requires an order of magnitude estimate of the widths. However, as we observed in section~\ref{sec:PS_constraints}, our theoretical widths are not useful for accurately predicting the physical width of the resulting gap, primarily because we do not know \textit{a priori} what value to assign to the threshold eccentricity $e_0$.

We could, however, calibrate $e_0$ against the simulations of section~\ref{sec:sims}. The model radial profiles shown in Fig.~\ref{fig:multi_sims} have a gap FWHM of around 20au. Table~\ref{tab:approx_widths} shows that the width of the `important' resonance, as defined in section~\ref{sec:sr_widths}, is given by $\frac{5}{7}\frac{E_2}{e_0} a_2$. Note that this depends on the initial eccentricity of the outermost planet only. Taking the values $E_2=0.05$ and $a_2=26\mathrm{au}$, as used in all simulations presented in the previous section, and equating the width to 20au gives $e_0\approx0.05$. This is interestingly close to $E_2$, and it may be tempting to assume that $e_0$ should in general be taken as $E_2$. However, this would imply that the physical width of the gap is in fact independent of $E_2$, which we know from the simulation discussed earlier in this subsection is not the case. Assuming instead that $e_0$ can always be taken as 0.05, we obtain

\begin{equation}\label{eqn:approxE2}
E_2\approx 0.07\frac{w}{a_2},
\end{equation}

where $w$ is the observed FWHM of the gap. The assumption that $e_0$ should be the same for all systems is not, however, realistic. The forced eccentricity close to the outermost planet -- i.e. at the inner edge of the disc -- is approximately equal to the planet's own eccentricity (\citealt{Murray1998}). Ideally, therefore, $e_0$ should not be smaller than $E_2$, since the presence of a gap depends on the fact that the planetesimal eccentricities near the secular resonances are higher than they are elsewhere -- but this is not guaranteed by equation~(\ref{eqn:approxE2}).

It is clear, then, that quantifying the resonance widths using Laplace-Lagrange theory is subject to a great deal of uncertainty. Using a higher-order theory able to estimate the maximum planetesimal eccentricity reached as a function of semi-major axis would be preferable, as this would eliminate the problem of trying to calculate the width of an infinite resonance. The ideal way to quantify the widths would be to run a grid of simulations covering the parameter space of planet eccentricities and evaluate them numerically, which is beyond the scope of this paper. Equation~(\ref{eqn:approxE2}) can nonetheless serve as a starting point for future $N$-body simulations.

\subsubsection{Wider Gaps from Sweeping Secular Resonances?}
\label{sec:migration}

\begin{figure*}
	\centering
	\includegraphics[width=0.95\textwidth]{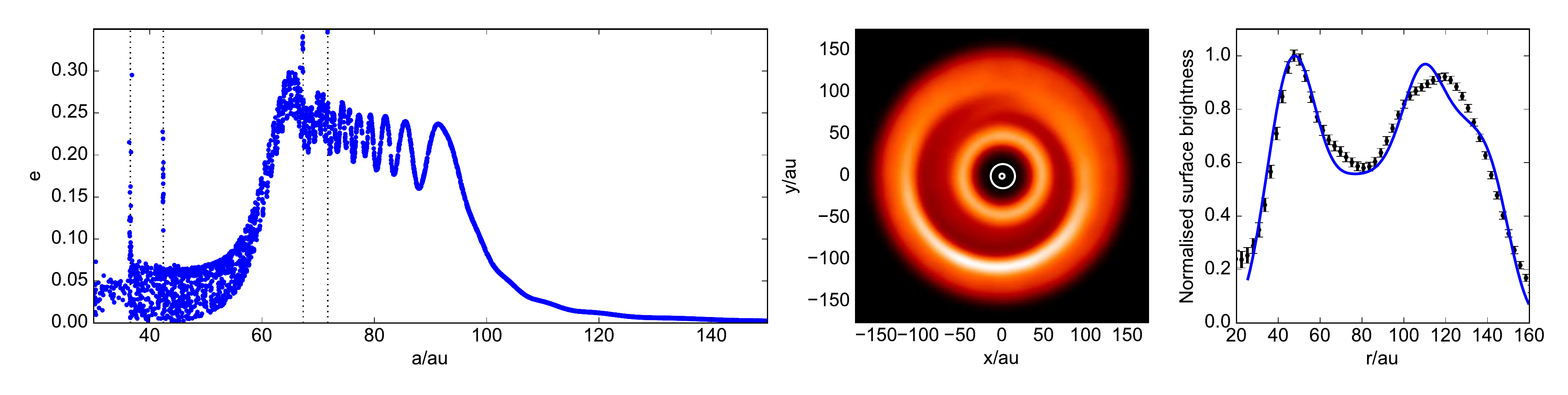}
	\caption{Summary of results of a simulation in which the outermost planet migrates outwards. Here $m_1=0.14M_{\mathrm{J}}$, $m_2=5M_{\mathrm{J}}$, $a_1=3\mathrm{au}$, and $a_2$ increases from 10.5 to 14.5au at a constant rate over 100Myr. Both planets initially had an eccentricity of 0.05. The left hand panel shows the eccentricity profile of the disc (along with the locations of the $4:1$, $5:1$, $10:1$ and $11:1$ mean motion resonances with the outer planet at its final position), the middle panel shows the resulting synthetic image (including collisional evolution), and the right hand panel shows the corresponding brightness profile and comparison with the ALMA data of \citet{Marino2018_107146}, all after 100Myr. }
	\label{fig:mig_sim}
\end{figure*}

Here we briefly investigate an alternative scenario which we expect to be able to produce a wider gap. Consider a system in which the secular resonances sculpting the disc are moving over time; we would then expect an increase in planetesimal eccentricities over the range of distances that any `important' resonances (i.e. wide resonances with short time-scales) sweep through. The resonances will move if either of the planets changes its semi-major axis, i.e. undergoes migration. Fig.~\ref{fig:mig_sim} summarises the results of a simulation with initial planetary eccentricities of 0.05, $m_1=0.14M_{\mathrm{J}}$, $m_2=5M_{\mathrm{J}}$, $a_1=3\mathrm{au}$, and $a_2$ changing from 10.5 to 14.5au at a constant rate over 100Myr. The `important' resonance here is the one at $r_2$, which increases from around 60 to 100au over 100Myr, and the resonance at $r_1$ is narrow. 

We forced the change in $a_2$ by introducing an additional tangential force on the outer planet. The final value of $a_2$ was chosen such that the $3:1$ mean motion resonance with the outer planet is at 30au, as we found in our simulations that if the $2:1$ or $3:1$ resonances are inside the disc, they clear the regions over which they sweep of planetesimals. For the collisional post-processing used to make the synthetic image and profile in Fig.~\ref{fig:mig_sim}, we took the planetesimal strength $Q_{\mathrm{D}}^{\star}$ to be $600\mathrm{J\,kg}^{-1}$, somewhat larger than the value used in section~\ref{sec:sims}, in order to ensure that the two peaks are of comparable brightness.

This simulation aims simply to exemplify the effect that migration can have on the depletion width, and here we do not give consideration to the mechanism responsible for the migration beyond noting that we would likely have to invoke interactions with planetesimals interior to the observed inner disc edge to explain it (e.g. \citealt{Levison2007_Migration}). 

It may in reality be difficult for such a massive planet (i.e. $5M_{\mathrm{J}}$ in the simulation we presented) to migrate significantly. We found that it was possible to obtain radial profiles similar to that shown in Fig.~\ref{fig:mig_sim} with less massive planets, but as this lengthens the secular time-scale, the initial planetary eccentricities had to be higher for planetesimal eccentricities to be excited to a level that gives a deep enough depletion. However, Fig.~\ref{fig:mig_sim} shows that the disc is already asymmetric, and increasing the planetary eccentricities has the disadvantage (for HD~107146) of making the spiral structure even more prominent.

The conclusion is that it is difficult within the secular resonance model to obtain a wide gap while maintaining a near-axisymmetric disc, and in the case of HD~107146 this issue is not alleviated by forcing the resonances to sweep through the disc. It may, however, be possible for wider depletions to form via secular resonance with migrating planets with less resulting asymmetry for much older systems, since then the migration time-scale could be much longer than the secular time-scale. In that case, the planetesimals in secular resonance would experience more secular precession time-scales before the resonance moves by a significant amount, and could thus reach large eccentricities even if the planet eccentricities were low.

\subsection{Further Model Applications}
\label{sec:other_systems}

Having considered in some detail the case of HD~107146, we now briefly discuss some other systems to which our model may be applicable. One strong candidate for such a system is HD~92945, a K-type star with an age of around 300Myr (\citealt{Plavchan2009_92945Age}). The debris disc around this star has a double-peaked radial surface brightness profile, with a depleted region at around 75au (\citealt{Golimowski2011_92945}; Marino et al. in preparation). Given that HD~107146 and HD~92945 have similar ages and, by chance, depletions at around the same location, the planetary parameters needed to recreate an HD~92945-like disc within our model would not differ greatly from those presented in section~\ref{sec:PS_constraints} for HD~107146. Under the assumption of disc truncation by the outer planet, the planets would have to be slightly further out, as the HD~92945 disc has its inner edge closer to 45 than 30au (\citealt{Golimowski2011_92945}). However, this assumption does not necessarily have to hold, which means that we can obtain an illustrative simulated profile to compare with data from one of the $N$-body simulations of section~\ref{sec:sims}, taking account of the fact that the HD~92945 disc is slightly narrower than that of HD~107146 by assigning zero mass to any test particles with semi-major axes outside some desired range. 

We find that using the results from simulation A (i.e. the parameters shown in Table~\ref{tab:exsys}) but keeping only test particles with semi-major axes between 45 and 130au in fact gives a profile close to the ALMA data of Marino et al. (in preparation), assuming an initially uniform surface density and no collisional evolution (i.e. stronger or larger planetesimals than we required for HD~107146). The requirement for no significant collisional depletion to have occurred is a result of the fact that in the case of HD~92945, the outer peak in the brightness profile is less bright than the inner, which contrasts with HD~107146 where the two are of comparable brightness. The gap in the HD~92945 disc is narrower than that in HD~107146, and as a result the simulations presented in this paper in fact match the profile of HD~92945 more closely; we expect to present a more detailed comparison between our model and ALMA observations of HD~92945 in a future paper.

Another potential application is to HD~38858, a Sun-like star which hosts a debris disc (\citealt{Beichman2006_38858Disc}) as well as one planet (\citealt{Mayor2011_38858Planet}; \citealt{Kennedy2015_RVlim}) detected using radial velocity observations. \citet{Kennedy2015_RVlim} compared several models for the disc with images from the \textit{Herschel} telescope, finding that the data were not consistent with a single narrow ring of debris, but could be well reproduced by either two planetesimal belts of width 10au located at around 55 and 130au or a single wide disc extending from around 30 to 200au. Thus, the detailed structure of the disc is not well constrained and it is possible that the reality is somewhere between these two models, with debris covering a wide range of distances and a depletion at some intermediate location. 

We can apply our model to determine where a second, undetected planet between the known planet and the disc could lie in mass--semi-major axis space such that the disc is depleted by the secular resonances of the system. If future high resolution observations of the HD~38858 disc were to reveal that there is indeed a depletion, this would motivate searching for a planet in those regions of parameter space (or conversely, if a planet were detected there, then the disc should be double-ringed, unless there are other planets present that change the dynamics of the disc). 

\begin{figure}
	\centering
    \hspace{-0.5cm}
	\includegraphics[width=1.\linewidth]{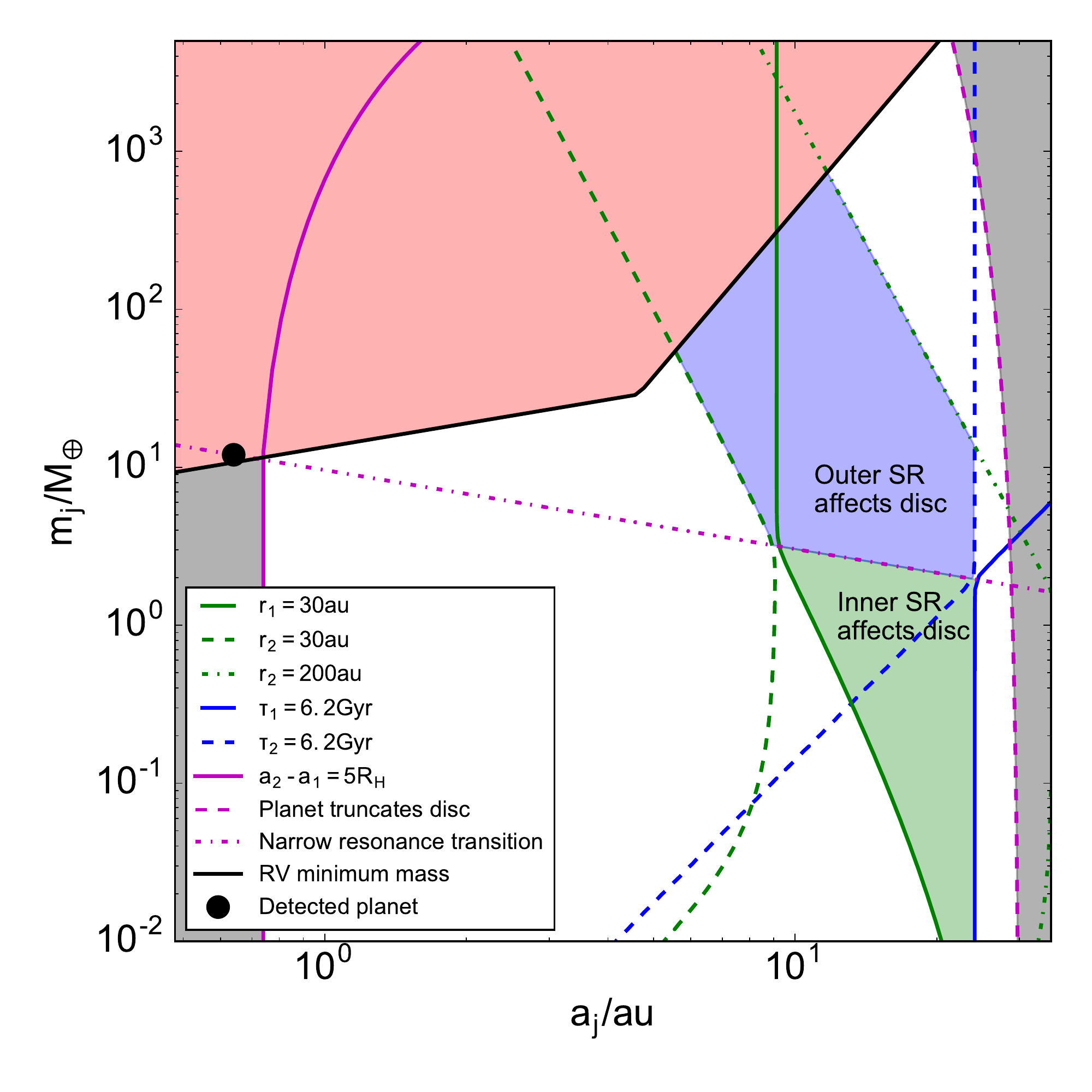}
	\caption{Plot showing the regions in mass--semi-major axis space where a planet orbiting HD 38858 exterior to the one known planet of this system would result in a configuration of secular resonances that affects the structure of the debris disc. The upper left corner is ruled out by radial velocity observations (\citealt{Kennedy2015_RVlim}). An outer planet in the lower left region would be unstable, and in the far right region it would truncate the inner edge of the disc. The two remaining highlighted regions show where each of the resonances in turn would be located between 30 and 200au, act on a time-scale less than the age of the system, and be wide enough to have an effect on the disc.}
	\label{fig:ps_38858}
\end{figure}

Fig.~\ref{fig:ps_38858} shows the results of this analysis. The quantities $a_1$ and $m_1$ are fixed at 0.64au and 12$M_\oplus$, the latter of which is the minimum possible mass of the known planet (\citealt{Kennedy2015_RVlim}); these parameters are shown by the black circle. Assuming the disc extends from 30 to 200au, the secular resonance at $r_i$ will be located in the disc provided the outer planet lies between the contours $r_i=30\mathrm{au}$ and $r_i=200\mathrm{au}$; these contours are shown in Fig.~\ref{fig:ps_38858}, with the exception of $r_1=200\mathrm{au}$, whose contour lies entirely within the disc. 

The resonance at $r_i$ will only have had time to excite eccentricities significantly if the corresponding secular time-scale $\tau_i$ is shorter than $t_{\mathrm{age}}$, the age of the system. This condition is satisfied if the outer planet is close enough in, and massive enough, that it lies to the left of the contour $\tau_i = t_{\mathrm{age}}$. Estimates of the age of HD~38858 vary widely; for illustrative purposes we adopt the `intermediate' value of 6.2Gyr reported by \citet{Wyatt2012_61Vir}. Note that the orientation of the $r_i$ and $\tau_i$ contours relative to each other is not the same as in section~\ref{sec:SR_theory}, for example curves of constant $r_1$ are no longer parallel to those of constant $\tau_1$. This is because we are now fixing the parameters of the inner planet rather than the outer; the results of Tables \ref{tab:limits} and \ref{tab:approx_ratios} still hold, but $a_2$ and $m_2$ are no longer constants, which changes the slopes of the curves.

Recall from section~\ref{sec:sr_widths} that above the line $\frac{m_2}{m_1}=\sqrt{\frac{a_1}{a_2}}$, the resonance at $r_1$ is very narrow, so that the disc will only be depleted significantly by that resonance if the hypothetical second planet lies below that line; the reverse is true for the resonance at $r_2$. The labelled regions in Fig.~\ref{fig:ps_38858} show where each of the secular resonances would affect the disc, based on the location, time-scale and width conditions outlined above. Note that radial velocity observations have ruled out the presence of planets in the upper left region (\citealt{Kennedy2015_RVlim}). The shaded regions at the left and right edges of the plot respectively show where the second planet would be unstable and where the disc would be perturbed by its resonance overlap zone, though these do not intersect the regions of interest.

A final system of possible interest is HD~131835, which has a disc composed of three rings (\citealt{Feldt2016_131835}) separated by depleted regions. The applicability of the secular resonance model is, however, less clear here, because the presence of gas in the system (\citealt{Moor2015_Gas}) means that gas-dust interactions (rather than only planet-planetesimal ones) could be important for explaining the disc's structure. If secular resonances were in fact responsible for the depleted regions, we would need to invoke more than two planets; we have showed that in the two-planet case, if the resonances are well-separated (as they would have to be to generate two separate depletions) then one of them will excite eccentricities only within a very narrow range of semi-major axes and thus will not produce any observable depletion.

\section{Conclusions}
\label{sec:conclusions}

We have investigated the formation of depleted regions in extended debris discs through eccentricity excitation by the exterior secular resonances of a two-planet system. We first used Laplace-Lagrange theory to understand how the locations, time-scales and widths of these resonances depend on the masses $m_j$, semi-major axes $a_j$ and eccentricities of the planets, finding in particular that one of the resonances will be too narrow to have any significant effect on the disc. These considerations were applied to the disc of HD 107146, which is known to have a depletion at $\sim$70--80au, in order to deduce the values of $m_1$ and $a_1$ that are expected to be able to give a disc structure similar to observations for a given choice of $m_2$ and $a_2$.

Guided by these parameter space constraints, we performed $N$-body simulations of five promising planetary configurations. We chose configurations spanning around one order of magnitude in semi-major axis $a_1$ and two in mass $m_1$; all simulations gave similar results. We found that planetesimal eccentricities are excited in the region where they are predicted to be, which leads to a clear depletion in the surface density there. The planets also induce a spiral structure in the disc just exterior to the depletion, on a scale smaller than the resolution of current observations.

We post-processed the $N$-body output using the method of \citet{Wyatt10_Collisions} to include the effect of collisional depletion, finding that collisions can deplete the inner part of the disc and the region where eccentricities are excited by secular resonances such that the relative heights of the peaks in the surface brightness profile, as well as the depth of the depletion, are comparable with the data of \citet{Marino2018_107146}. While our simulated brightness profiles do not agree with the data within the observational uncertainties, the simulated discs are qualitatively similar to HD~107146. 

Based on our simulations, we concluded that the secular resonance model is best suited to systems with asymmetric gaps that are deeper and wider on one side, because the inner ring of the disc always has an offset relative to the outer. Finally, we found that it is possible to obtain a gap whose width is comparable with that of HD~107146 -- though at the expense of more prominent spiral structure -- by forcing one of the planets to migrate over time, such that the secular resonances sweep over a range of distances.

\section*{Acknowledgements}

BY acknowledges the support of an STFC studentship. GMK is supported by the Royal Society as a Royal Society University Research Fellow. We thank Mark Wyatt for helpful discussions about the paper and for providing a code for the calculation of collision rates. We also thank Alessandro Morbidelli for his constructive review which improved the clarity of the paper.




\bibliographystyle{mnras}
\bibliography{refs}

\begin{thebibliography}{}
\makeatletter
\relax
\def\mn@urlcharsother{\let\do\@makeother \do\$\do\&\do\#\do\^\do\_\do\%\do\~}
\def\mn@doi{\begingroup\mn@urlcharsother \@ifnextchar [ {\mn@doi@}
  {\mn@doi@[]}}
\def\mn@doi@[#1]#2{\def\@tempa{#1}\ifx\@tempa\@empty \href
  {http://dx.doi.org/#2} {doi:#2}\else \href {http://dx.doi.org/#2} {#1}\fi
  \endgroup}
\def\mn@eprint#1#2{\mn@eprint@#1:#2::\@nil}
\def\mn@eprint@arXiv#1{\href {http://arxiv.org/abs/#1} {{\tt arXiv:#1}}}
\def\mn@eprint@dblp#1{\href {http://dblp.uni-trier.de/rec/bibtex/#1.xml}
  {dblp:#1}}
\def\mn@eprint@#1:#2:#3:#4\@nil{\def\@tempa {#1}\def\@tempb {#2}\def\@tempc
  {#3}\ifx \@tempc \@empty \let \@tempc \@tempb \let \@tempb \@tempa \fi \ifx
  \@tempb \@empty \def\@tempb {arXiv}\fi \@ifundefined
  {mn@eprint@\@tempb}{\@tempb:\@tempc}{\expandafter \expandafter \csname
  mn@eprint@\@tempb\endcsname \expandafter{\@tempc}}}

\bibitem[\protect\citeauthoryear{{Andrews}, {Wilner}, {Hughes}, {Qi}  \&
  {Dullemond}}{{Andrews} et~al.}{2009}]{Andrews2009_PPDisks}
{Andrews} S.~M.,  {Wilner} D.~J.,  {Hughes} A.~M.,  {Qi} C.,   {Dullemond}
  C.~P.,  2009, \mn@doi [\apj] {10.1088/0004-637X/700/2/1502}, \href
  {http://adsabs.harvard.edu/abs/2009ApJ...700.1502A} {700, 1502}

\bibitem[\protect\citeauthoryear{{Apai} et~al.,}{{Apai}
  et~al.}{2008}]{Apai2008_DirectImg}
{Apai} D.,  et~al., 2008, \mn@doi [\apj] {10.1086/524191}, \href
  {http://adsabs.harvard.edu/abs/2008ApJ...672.1196A} {672, 1196}

\bibitem[\protect\citeauthoryear{{Backman} \& {Paresce}}{{Backman} \&
  {Paresce}}{1993}]{Backman1993_Circumstellar}
{Backman} D.~E.,  {Paresce} F.,  1993, in {Levy} E.~H.,  {Lunine} J.~I.,  eds,
  Protostars and Planets III. pp 1253--1304

\bibitem[\protect\citeauthoryear{{Beichman} et~al.,}{{Beichman}
  et~al.}{2006}]{Beichman2006_38858Disc}
{Beichman} C.~A.,  et~al., 2006, \mn@doi [\apj] {10.1086/508449}, \href
  {http://adsabs.harvard.edu/abs/2006ApJ...652.1674B} {652, 1674}

\bibitem[\protect\citeauthoryear{{Booth} et~al.,}{{Booth}
  et~al.}{2016}]{Booth2016_8799}
{Booth} M.,  et~al., 2016, \mn@doi [\mnras] {10.1093/mnrasl/slw040}, \href
  {http://adsabs.harvard.edu/abs/2016MNRAS.460L..10B} {460, L10}

\bibitem[\protect\citeauthoryear{{Davies}, {Adams}, {Armitage}, {Chambers},
  {Ford}, {Morbidelli}, {Raymond}  \& {Veras}}{{Davies}
  et~al.}{2014}]{Davies2014_Hill}
{Davies} M.~B.,  {Adams} F.~C.,  {Armitage} P.,  {Chambers} J.,  {Ford} E.,
  {Morbidelli} A.,  {Raymond} S.~N.,   {Veras} D.,  2014, \mn@doi [Protostars
  and Planets VI] {10.2458/azu_uapress_9780816531240-ch034}, \href
  {http://adsabs.harvard.edu/abs/2014prpl.conf..787D} {pp 787--808}

\bibitem[\protect\citeauthoryear{{Dohnanyi}}{{Dohnanyi}}{1968}]{Dohnanyi68_CCindex}
{Dohnanyi} J.~S.,  1968, in {Kresak} L.,  {Millman} P.~M.,  eds,  IAU Symposium
  Vol. 33, Physics and Dynamics of Meteors. p.~486

\bibitem[\protect\citeauthoryear{{Dominik} \& {Decin}}{{Dominik} \&
  {Decin}}{2003}]{Dominik2003_CC}
{Dominik} C.,  {Decin} G.,  2003, \mn@doi [\apj] {10.1086/379169}, \href
  {http://adsabs.harvard.edu/abs/2003ApJ...598..626D} {598, 626}

\bibitem[\protect\citeauthoryear{{Feldt} et~al.,}{{Feldt}
  et~al.}{2017}]{Feldt2016_131835}
{Feldt} M.,  et~al., 2017, \mn@doi [\aap] {10.1051/0004-6361/201629261}, \href
  {http://adsabs.harvard.edu/abs/2017A%26A...601A...7F} {601, A7}

\bibitem[\protect\citeauthoryear{{Golimowski} et~al.,}{{Golimowski}
  et~al.}{2011}]{Golimowski2011_92945}
{Golimowski} D.~A.,  et~al., 2011, \mn@doi [\aj] {10.1088/0004-6256/142/1/30},
  \href {http://adsabs.harvard.edu/abs/2011AJ....142...30G} {142, 30}

\bibitem[\protect\citeauthoryear{{Kennedy} et~al.,}{{Kennedy}
  et~al.}{2015}]{Kennedy2015_RVlim}
{Kennedy} G.~M.,  et~al., 2015, \mn@doi [\mnras] {10.1093/mnras/stv511}, \href
  {http://adsabs.harvard.edu/abs/2015MNRAS.449.3121K} {449, 3121}

\bibitem[\protect\citeauthoryear{{Lagrange} et~al.,}{{Lagrange}
  et~al.}{2010}]{Lagrange2010_betaPicImage}
{Lagrange} A.-M.,  et~al., 2010, \mn@doi [Science] {10.1126/science.1187187},
  \href {http://adsabs.harvard.edu/abs/2010Sci...329...57L} {329, 57}

\bibitem[\protect\citeauthoryear{{Levison}, {Morbidelli}, {Gomes}  \&
  {Backman}}{{Levison} et~al.}{2007}]{Levison2007_Migration}
{Levison} H.~F.,  {Morbidelli} A.,  {Gomes} R.,   {Backman} D.,  2007,
  Protostars and Planets V, \href
  {http://adsabs.harvard.edu/abs/2007prpl.conf..669L} {pp 669--684}

\bibitem[\protect\citeauthoryear{{Malhotra}}{{Malhotra}}{1998}]{Malhotra98_Resonances}
{Malhotra} R.,  1998, in {Lazzaro} D.,  {Vieira Martins} R.,  {Ferraz-Mello}
  S.,   {Fernandez} J.,  eds,  Astronomical Society of the Pacific Conference
  Series Vol. 149, Solar System Formation and Evolution. p.~37

\bibitem[\protect\citeauthoryear{{Marino}, {Wyatt}, {Kennedy}, {Holland},
  {Matr{\`a}}, {Shannon}  \& {Ivison}}{{Marino}
  et~al.}{2017}]{Marino2017_61Vir}
{Marino} S.,  {Wyatt} M.~C.,  {Kennedy} G.~M.,  {Holland} W.,  {Matr{\`a}} L.,
  {Shannon} A.,   {Ivison} R.~J.,  2017, \mn@doi [\mnras]
  {10.1093/mnras/stx1102}, \href
  {http://adsabs.harvard.edu/abs/2017MNRAS.469.3518M} {469, 3518}

\bibitem[\protect\citeauthoryear{{Marino} et~al.,}{{Marino}
  et~al.}{2018}]{Marino2018_107146}
{Marino} S.,  et~al., 2018, preprint, \href
  {http://adsabs.harvard.edu/abs/2018arXiv180501915M} {} (\mn@eprint {arXiv}
  {1805.01915})

\bibitem[\protect\citeauthoryear{{Marois}, {Macintosh}, {Barman}, {Zuckerman},
  {Song}, {Patience}, {Lafreni{\`e}re}  \& {Doyon}}{{Marois}
  et~al.}{2008}]{Marois2008_8799}
{Marois} C.,  {Macintosh} B.,  {Barman} T.,  {Zuckerman} B.,  {Song} I.,
  {Patience} J.,  {Lafreni{\`e}re} D.,   {Doyon} R.,  2008, \mn@doi [Science]
  {10.1126/science.1166585}, \href
  {http://adsabs.harvard.edu/abs/2008Sci...322.1348M} {322, 1348}

\bibitem[\protect\citeauthoryear{{Marois}, {Zuckerman}, {Konopacky},
  {Macintosh}  \& {Barman}}{{Marois} et~al.}{2010}]{Marois2010_8799}
{Marois} C.,  {Zuckerman} B.,  {Konopacky} Q.~M.,  {Macintosh} B.,   {Barman}
  T.,  2010, \mn@doi [\nat] {10.1038/nature09684}, \href
  {http://adsabs.harvard.edu/abs/2010Natur.468.1080M} {468, 1080}

\bibitem[\protect\citeauthoryear{{Mayor} et~al.,}{{Mayor}
  et~al.}{2011}]{Mayor2011_38858Planet}
{Mayor} M.,  et~al., 2011, preprint, \href
  {http://adsabs.harvard.edu/abs/2011arXiv1109.2497M} {} (\mn@eprint {arXiv}
  {1109.2497})

\bibitem[\protect\citeauthoryear{{Milani} \& {Knezevic}}{{Milani} \&
  {Knezevic}}{1990}]{Milani90_SecRes}
{Milani} A.,  {Knezevic} Z.,  1990, \mn@doi [Celestial Mechanics and Dynamical
  Astronomy] {10.1007/BF00049444}, \href
  {http://adsabs.harvard.edu/abs/1990CeMDA..49..347M} {49, 347}

\bibitem[\protect\citeauthoryear{{Mo{\'o}r} et~al.,}{{Mo{\'o}r}
  et~al.}{2015}]{Moor2015_Gas}
{Mo{\'o}r} A.,  et~al., 2015, \mn@doi [\apj] {10.1088/0004-637X/814/1/42},
  \href {http://adsabs.harvard.edu/abs/2015ApJ...814...42M} {814, 42}

\bibitem[\protect\citeauthoryear{{Moro-Mart{\'{\i}}n}
  et~al.,}{{Moro-Mart{\'{\i}}n} et~al.}{2007}]{MoroMartin08_38529}
{Moro-Mart{\'{\i}}n} A.,  et~al., 2007, \mn@doi [\apj] {10.1086/521093}, \href
  {http://adsabs.harvard.edu/abs/2007ApJ...668.1165M} {668, 1165}

\bibitem[\protect\citeauthoryear{{Mouillet}, {Larwood}, {Papaloizou}  \&
  {Lagrange}}{{Mouillet} et~al.}{1997}]{Mouillet97_betaPicPred}
{Mouillet} D.,  {Larwood} J.~D.,  {Papaloizou} J.~C.~B.,   {Lagrange} A.~M.,
  1997, \mn@doi [\mnras] {10.1093/mnras/292.4.896}, \href
  {http://adsabs.harvard.edu/abs/1997MNRAS.292..896M} {292, 896}

\bibitem[\protect\citeauthoryear{{Murray} \& {Dermott}}{{Murray} \&
  {Dermott}}{1998}]{Murray1998}
{Murray} C.~D.,  {Dermott} S.~F.,  1998, Solar System Dynamics.
Cambridge University Press

\bibitem[\protect\citeauthoryear{{Pearce} \& {Wyatt}}{{Pearce} \&
  {Wyatt}}{2015}]{Pearce15_EccPlanet}
{Pearce} T.~D.,  {Wyatt} M.~C.,  2015, \mn@doi [\mnras]
  {10.1093/mnras/stv1847}, \href
  {http://adsabs.harvard.edu/abs/2015MNRAS.453.3329P} {453, 3329}

\bibitem[\protect\citeauthoryear{{Plavchan}, {Werner}, {Chen}, {Stapelfeldt},
  {Su}, {Stauffer}  \& {Song}}{{Plavchan} et~al.}{2009}]{Plavchan2009_92945Age}
{Plavchan} P.,  {Werner} M.~W.,  {Chen} C.~H.,  {Stapelfeldt} K.~R.,  {Su}
  K.~Y.~L.,  {Stauffer} J.~R.,   {Song} I.,  2009, \mn@doi [\apj]
  {10.1088/0004-637X/698/2/1068}, \href
  {http://adsabs.harvard.edu/abs/2009ApJ...698.1068P} {698, 1068}

\bibitem[\protect\citeauthoryear{{Rein} \& {Liu}}{{Rein} \&
  {Liu}}{2012}]{Rein2012_Rebound}
{Rein} H.,  {Liu} S.-F.,  2012, \mn@doi [\aap] {10.1051/0004-6361/201118085},
  \href {http://adsabs.harvard.edu/abs/2012A%26A...537A.128R} {537, A128}

\bibitem[\protect\citeauthoryear{{Rein} \& {Tamayo}}{{Rein} \&
  {Tamayo}}{2015}]{Rein2015_whfast}
{Rein} H.,  {Tamayo} D.,  2015, \mn@doi [\mnras] {10.1093/mnras/stv1257}, \href
  {http://adsabs.harvard.edu/abs/2015MNRAS.452..376R} {452, 376}

\bibitem[\protect\citeauthoryear{{Ricci}, {Carpenter}, {Fu}, {Hughes}, {Corder}
   \& {Isella}}{{Ricci} et~al.}{2015}]{Ricci15_AlmaObs}
{Ricci} L.,  {Carpenter} J.~M.,  {Fu} B.,  {Hughes} A.~M.,  {Corder} S.,
  {Isella} A.,  2015, \mn@doi [\apj] {10.1088/0004-637X/798/2/124}, \href
  {http://adsabs.harvard.edu/abs/2015ApJ...798..124R} {798, 124}

\bibitem[\protect\citeauthoryear{{Schneider}, {Dedieu}, {Le Sidaner}, {Savalle}
   \& {Zolotukhin}}{{Schneider} et~al.}{2011}]{Schneider2011_Cat}
{Schneider} J.,  {Dedieu} C.,  {Le Sidaner} P.,  {Savalle} R.,   {Zolotukhin}
  I.,  2011, \mn@doi [\aap] {10.1051/0004-6361/201116713}, \href
  {http://adsabs.harvard.edu/abs/2011A%26A...532A..79S} {532, A79}

\bibitem[\protect\citeauthoryear{{Watson}, {Littlefair}, {Diamond}, {Collier
  Cameron}, {Fitzsimmons}, {Simpson}, {Moulds}  \& {Pollacco}}{{Watson}
  et~al.}{2011}]{Watson11_MassRad}
{Watson} C.~A.,  {Littlefair} S.~P.,  {Diamond} C.,  {Collier Cameron} A.,
  {Fitzsimmons} A.,  {Simpson} E.,  {Moulds} V.,   {Pollacco} D.,  2011,
  \mn@doi [\mnras] {10.1111/j.1745-3933.2011.01036.x}, \href
  {http://adsabs.harvard.edu/abs/2011MNRAS.413L..71W} {413, L71}

\bibitem[\protect\citeauthoryear{{Williams}, {Najita}, {Liu}, {Bottinelli},
  {Carpenter}, {Hillenbrand}, {Meyer}  \& {Soderblom}}{{Williams}
  et~al.}{2004}]{Williams04_Age}
{Williams} J.~P.,  {Najita} J.,  {Liu} M.~C.,  {Bottinelli} S.,  {Carpenter}
  J.~M.,  {Hillenbrand} L.~A.,  {Meyer} M.~R.,   {Soderblom} D.~R.,  2004,
  \mn@doi [\apj] {10.1086/381721}, \href
  {http://adsabs.harvard.edu/abs/2004ApJ...604..414W} {604, 414}

\bibitem[\protect\citeauthoryear{{Wisdom}}{{Wisdom}}{1980}]{Wisdom80_ResOverlap}
{Wisdom} J.,  1980, \mn@doi [\aj] {10.1086/112778}, \href
  {http://adsabs.harvard.edu/abs/1980AJ.....85.1122W} {85, 1122}

\bibitem[\protect\citeauthoryear{{Wyatt}}{{Wyatt}}{2008}]{Wyatt08_Review}
{Wyatt} M.~C.,  2008, \mn@doi [\araa] {10.1146/annurev.astro.45.051806.110525},
  \href {http://adsabs.harvard.edu/abs/2008ARA%26A..46..339W} {46, 339}

\bibitem[\protect\citeauthoryear{{Wyatt}, {Booth}, {Payne}  \&
  {Churcher}}{{Wyatt} et~al.}{2010}]{Wyatt10_Collisions}
{Wyatt} M.~C.,  {Booth} M.,  {Payne} M.~J.,   {Churcher} L.~J.,  2010, \mn@doi
  [\mnras] {10.1111/j.1365-2966.2009.15930.x}, \href
  {http://adsabs.harvard.edu/abs/2010MNRAS.402..657W} {402, 657}

\bibitem[\protect\citeauthoryear{{Wyatt} et~al.,}{{Wyatt}
  et~al.}{2012}]{Wyatt2012_61Vir}
{Wyatt} M.~C.,  et~al., 2012, \mn@doi [\mnras]
  {10.1111/j.1365-2966.2012.21298.x}, \href
  {http://adsabs.harvard.edu/abs/2012MNRAS.424.1206W} {424, 1206}

\bibitem[\protect\citeauthoryear{{Zheng}, {Lin}, {Kouwenhoven}, {Mao}  \&
  {Zhang}}{{Zheng} et~al.}{2017}]{Zheng2017_GasSR}
{Zheng} X.,  {Lin} D.~N.~C.,  {Kouwenhoven} M.~B.~N.,  {Mao} S.,   {Zhang} X.,
  2017, \mn@doi [\apj] {10.3847/1538-4357/aa8ef3}, \href
  {http://adsabs.harvard.edu/abs/2017ApJ...849...98Z} {849, 98}

\makeatother
\end{thebibliography}







\bsp	
\label{lastpage}
\end{document}